\documentclass[prd,aps,preprintnumbers,nofootinbib,superscriptaddress,notitlepage,floatfix,11pt]{revtex4}
\usepackage{stmaryrd}
\usepackage{epsfig}
\usepackage{amsfonts}
\usepackage{amsmath}
\usepackage{graphicx}
\usepackage{color}
\usepackage{amssymb,amsmath,psfrag,slashed,graphicx}
\usepackage[normalem]{ulem}
\usepackage{subfigure}
\usepackage{array}

\def\slash{\!\!\!/}
\newcommand{\nn}{\nonumber}

\allowdisplaybreaks

\begin{document}

\title{Doubly Charged Higgs Production at Future $ep$ Colliders}

\author{Xing-Hua Yang~\footnote{yangxinghua@sdut.edu.cn}}
\affiliation{School of Physics and Optoelectronic Engineering, Shandong University of Technology, Zibo, Shandong 255000, China}

\author{Zhong-Juan Yang~\footnote{sps\_yangzj@ujn.edu.cn}}
\affiliation{School of Physics and Technology, University of Jinan, Jinan, Shandong 250022, China}

\begin{abstract}

The Higgs sector of the standard model can be extended by introducing an $SU(2)_L$ Higgs triplet $\Delta$ to generate the tiny neutrino masses in the framework of type-II seesaw mechanism.
In this paper, we study the pair production of the introduced Higgs triplet at future $e^{-}p$ colliders.
The corresponding production cross sections via vector boson fusion process at FCC-ep and ILC$\otimes$FCC are predicted, where the production of a pair of doubly charged Higgs is found to be dominant and then used to investigate the collider phenomenology of the Higgs triplet.
Depending on the size of the Higgs triplet vacuum expectation value, the doubly charged Higgs may decay into a pair of same-sign charged leptons or a pair of same-sign $W$ bosons.
In order to explore the discovery potential of the doubly charged Higgs at future $e^{-}p$ colliders, we discuss these two decay scenarios in detail and show respectively the detection sensitivity on the mass of the doubly charged Higgs.

\end{abstract}

\maketitle

\section{Introduction}\label{sec1}

The experimental observation of neutrino oscillations has shown that neutrinos are massive and lepton flavors are mixed, which clearly indicates the existence of new physics beyond the standard model.
In order to accommodate the tiny neutrino masses, the natural way is to introduce the unique Weinberg dimension-five operator $LLHH/\Lambda$~\cite{Weinberg:1979sa}, where $L$ and $H$ denote respectively the lepton and Higgs doublet, and $\Lambda$ is the cut-off scale of new physics.
After spontaneous gauge symmetry breaking, the Weinberg dimension-five operator gives rise to Majorana neutrino masses $m_{\nu} \sim \langle H \rangle^2/\Lambda$ with $\langle H \rangle$ being the vacuum expectation value (vev) of the Higgs doublet, and thus the smallness of the neutrino masses can be ascribed to the existence of a large new scale $\Lambda$.
At tree-level, there are only three generic ways to obtain the Weinberg dimension-five operator, namely, type-I~\cite{Minkowski:1977sc,Yanagida:1979ss,Gell-Mann:1979ss,Glashow:1979ss,Mohapatra:1979ia}, type-II~~\cite{Konetschny:1977bn,Magg:1980ut,Schechter:1980gr,Lazarides:1980nt,Mohapatra:1980yp} and type-III~~\cite{Foot:1988aq,Ma:1998dn} seesaw mechanisms, where three $SU(2)_{L}$ singlet right-handed neutrinos, an $SU(2)_{L}$ Higgs triplet and three $SU(2)_{L}$ triplet fermions are added to the standard model, respectively.
The key point to test the seesaw mechanisms is to search for the existence of the introduced heavy states.
Since all the three seesaw mechanisms violate the lepton-number in their unique ways, we can probe the production signal of the relevant heavy particles via the lepton-number violating processes at ongoing and forthcoming experiments, if the mass scale of the heavy particles lies around TeV.
In this paper, we investigate the collider phenomenology of the Higgs triplet introduced in type-II seesaw mechanism at future $e^{-}p$ colliders.

A typical feature of the type-II seesaw mechanism is that the introduced Higgs triplet can be produced directly through gauge interactions with the electroweak bosons.
In the framework of type-II seesaw mechanism, there are seven physical Higgs bosons (i.e., $H^{++}$, $H^{--}$, $H^{+}$, $H^{-}$, $H^{0}$, $h^{0}$ and $A^{0}$), and the searches for the new triplet scalars have been studied extensively at various collider experiments, see Ref.~\cite{Cai:2017mow} for recent reviews.
At hadron colliders, the new triplet scalars are mainly produced in pair, since their single production and associated production with gauge bosons are highly suppressed by the small Higgs triplet vev.
Specifically, the most relevant production channels are the Drell-Yan processes via $s$-channel $\gamma^\ast/Z^\ast$ or $W^\ast$ exchange~\cite{Barger:1982cy,Gunion:1989in,Dion:1998pw,Muhlleitner:2003me,Akeroyd:2005gt,Perez:2008ha,delAguila:2008cj,Arhrib:2011uy,Li:2018jns,Du:2018eaw,Primulando:2019evb}.
A pair of triplet scalars can also be produced via vector boson fusion process~\cite{Dutta:2014dba,Bambhaniya:2015wna}, among which the charged Higgs pair production via photon fusion process is of special interest due to the contribution from collinear photon that includes both elastic and inelastic processes~\cite{Drees:1994zx,Han:2007bk,Babu:2016rcr}.
In addition, the pair production of the triplet scalars via gluon fusion process is found to be sub-leading with respect to the Drell-Yan process~\cite{Hessler:2014ssa,Nemevsek:2016enw}.
At $e^{+}e^{-}$ colliders, the most widely studied mode for the triplet scalars is the pair production via $s$-channel $\gamma^\ast/Z^\ast$ exchange~\cite{Komamiya:1988rs,Gunion:1989ci,Frank:1995ex,Ghosh:1996jg}.
At $e^{-}p$ colliders, the single production of the triplet scalars with signal rate directly proportional to the Yukuwa coupling between
the lepton doublet and the Higgs triplet has been discussed in some earlier studies~\cite{Accomando:1993ar,Yue:2007ym,Dev:2019hev}.
Complementary to the previous studies, in this work, we investigate the production and decay of the triplet scalars via vector boson fusion process at future $e^{-}p$ colliders, such as FCC-ep~\cite{Bruning:2260408} and ILC$\otimes$FCC~\cite{Acar:2016rde}.
Since the single production of the triplet scalars via vector boson fusion process is also highly suppressed by the small Higgs triplet vev, we focus on the production of a pair of triplet scalars.
The dominant production channel considered here is the pair production of the doubly charged Higgs, which may decay into same-sign dileptons ($\ell^{\pm}\ell^{\pm}$) or same-sign dibosons ($W^{\pm}W^{\pm}$) depending on the size of the Higgs triplet vev.
In order to explore the discovery potential of the doubly charged Higgs at future $e^{-}p$ colliders, we discuss these two decay scenarios respectively.

The rest of the paper is organized as follows. The main properties of the type-II seesaw model are briefly reviewed in Section~\ref{sec2}, and the various constraints on the model parameters are summarized in Section~\ref{sec3}.
In Section~\ref{sec4}, the dominant production channels of the triplet scalars via vector boson fusion process and their decay properties are studied.
The signal observability at future $e^{-}p$ colliders for both the lepton decay mode and the gauge boson decay mode is discussed in Section~\ref{sec5}.
Finally, we summarize in Section~\ref{sec6}.

\section{The model}\label{sec2}

In the type-II seesaw model, the Higgs sector is composed of the standard model Higgs doublet $H$ with hypercharge $Y_{H}=1$ and an $SU(2)_{L}$ Higgs triplet $\Delta$ with hypercharge $Y_{\Delta}=2$, which can be expressed in the matrix forms as
\begin{eqnarray}
\label{1}
H = \left(\begin{array}{c} \phi^{+} \\ \phi^{0} \end{array}\right)  \; , \quad
\Delta = \left(\begin{array}{cc} \delta^{+}/\sqrt{2} & \delta^{++} \\ \delta^{0} & -\delta^{+}/\sqrt{2}\end{array}\right) \; ,
\end{eqnarray}
where $\phi^{+}$, $\phi^{0}$, $\delta^{++}$, $\delta^{+}$ and $\delta^{0}$ are all complex scalar fields, and therefore there are total 10 degrees of freedom in the Higgs sector.
The most general gauge-invariant Lagrangian relevant for the Higgs sector can be given by
\begin{eqnarray}
\label{2}
{\cal L}_{\rm type-II} = (D_{\mu}H)^{\dagger}(D^{\mu}H) + {\rm Tr}\left[(D_{\mu}\Delta)^{\dagger}(D^{\mu}\Delta)\right] -V(H,\Delta)+{\cal L}_{\rm Yukawa} \; .
\end{eqnarray}
Here the covariant derivatives are defined as
\begin{eqnarray}
\label{3}
D_{\mu}H &\equiv& \partial_{\mu}H + i g \tau^{k} W_{\mu}^{k}H + i g^{\prime} \frac{Y_{H}}{2} B_{\mu}H \; , \nn \\
D_{\mu}\Delta &\equiv& \partial_{\mu}\Delta + i g \left[\tau^{k} W_{\mu}^{k},\Delta\right] + i g^{\prime} \frac{Y_{\Delta}}{2} B_{\mu}\Delta \; ,
\end{eqnarray}
where $W_{\mu}^{k}$ ($k=1,2,3$) and $B_{\mu}$ are the $SU(2)_{L}$ and $U(1)_Y$ gauge fields, respectively. $g$ and $g^{\prime}$ are the corresponding gauge couplings, and $\tau^{k}=\sigma^{k}/2$ ($k=1,2,3$) represents the $SU(2)_{L}$ generator with $\sigma^{k}$ being the Pauli matrices.

There are seven physical massive Higgs bosons in the model, namely, doubly charged $H^{++}$ and $H^{--}$, singly charged $H^{+}$ and $H^{-}$, CP-even neutral $H^{0}$ and $h^{0}$, and CP-odd neutral $A^{0}$, where $h^{0}$ is marked as the SM-like Higgs boson and the rest of the Higgs states are all $\Delta$-like.
To derive the Higgs mass spectrum, a detailed study on the Higgs potential should be given.
For simplicity, we just focus on the minimal Higgs potential
\begin{eqnarray}
\label{4}
V(H, \Delta)=-m_{H}^{2} H^{\dagger} H+\frac{\lambda}{4}\left(H^{\dagger} H\right)^{2}
+M_{\Delta}^{2} {\rm Tr} \left(\Delta^{\dagger} \Delta\right)
+\left(\mu H^{T} i \sigma^{2} \Delta^{\dagger} H+{\rm h.c.}\right) \; .
\end{eqnarray}
Here $m_{H}$ and $M_{\Delta}$ are the mass parameters, $\lambda$ is a dimensionless coupling. It is worth mentioning that the last term of Eq.~(\ref{4}) denotes the mixing between the Higgs doublet and triplet via a dimensional parameter $\mu$.
When the neutral components of $H$ and $\Delta$ acquire their vevs
\begin{eqnarray}
\label{5}
\langle H \rangle= \left(\begin{array}{c} 0 \\ v_{H}/\sqrt{2} \end{array}\right)   \quad {\rm and}   \quad
\langle \Delta \rangle = \left(\begin{array}{cc} 0 & 0 \\ v_{\Delta}/\sqrt{2} & 0\end{array}\right) \; ,
\end{eqnarray}
the gauge symmetry is spontaneously broken down.
After minimizing the minimal Higgs potential,
we can easily obtain
\begin{eqnarray}
\label{6}
v_{H} = \sqrt{\frac{4m_{H}^{2}M_{\Delta}^{2}}{\lambda M_{\Delta}^{2}-4\mu^{2}}} \; ,  \quad
v_{\Delta} = \frac{\mu v_{H}^{2}}{\sqrt{2}M_{\Delta}^{2}} \; ,
\end{eqnarray}
with $\sqrt{v_{H}^{2}+v_{\Delta}^{2}}\approx 246~{\rm GeV}$, where $\lambda M_{\Delta}^{2}-4\mu^{2}>0$ has been assumed.
Note that the Higgs triplet vev $v_{\Delta}$ contributes to the electroweak gauge boson masses, and hence the $\rho$-parameter at tree-level.
According to Eq.~(\ref{5}), the Higgs doublet and triplet can be redefined respectively as
\begin{eqnarray}
\label{7}
H = \left(\begin{array}{c} \phi^{+} \\ (v_{H} + \xi + i\chi)/\sqrt{2} \end{array}\right)  \; ,  \quad
\Delta = \left(\begin{array}{cc} \delta^{+}/\sqrt{2} & \delta^{++} \\ (v_{\Delta} + \zeta + i\eta)/\sqrt{2} & ~-\delta^{+}/\sqrt{2}~\end{array}\right) \; .
\end{eqnarray}
Here $\xi$, $\chi$, $\zeta$ and $\eta$ are real scalar fields with zero vevs.
Plugging Eq.~(\ref{7}) into Eq.~(\ref{4}), we have
\begin{align}
\label{8}
V(H, \Delta)\supset& M_{\Delta}^{2}\delta^{++}\delta^{--} + \sqrt{2}\mu v_{\Delta}\phi^{+}\phi^{-}
-\mu v_{H}\left(\phi^{+}\delta^{-}+\phi^{-}\delta^{+}\right)+M_{\Delta}^{2}\delta^{+}\delta^{-} \nn \\
&+\frac{1}{4}\lambda v_{H}^{2}\xi^{2}-\sqrt{2}\mu v_{H}\xi\zeta +\frac{1}{2}M_{\Delta}^{2}\zeta^{2}
+\sqrt{2}\mu v_{\Delta}\chi^{2} - \sqrt{2}\mu v_{H}\chi\eta+\frac{1}{2}M_{\Delta}^{2}\eta^{2} \; ,
\end{align}
where only the Higgs mass terms are retained and the relationships in Eq.~(\ref{6}) have been used.
Obviously, the doubly charged $\delta^{\pm\pm}$ are their mass eigenstates, while the singly charged $(\phi^{\pm},~\delta^{\pm})$, the CP-even neutral $(\xi,~\zeta)$ and the CP-odd neutral $(\chi,~\eta)$ mix with each other, respectively.
Eq.~(\ref{8}) can be further rewritten as
\begin{align}
\label{9}
V(H, \Delta)\supset&
M_{\Delta}^{2}\delta^{++}\delta^{--}
+ \left(\begin{array}{cc} \phi^{+} & \delta^{+} \end{array}\right)M_{\pm}^{2} \left(\begin{array}{c} \phi^{-} \\ \delta^{-} \end{array}\right) \nn \\
&+ \left(\begin{array}{cc} \xi & \zeta \end{array}\right)\frac{1}{2}M_{\rm even}^{2} \left(\begin{array}{c} \xi \\ \zeta \end{array}\right)
+  \left(\begin{array}{cc} \chi & \eta \end{array}\right)\frac{1}{2}M_{\rm odd}^{2} \left(\begin{array}{c} \chi \\ \eta \end{array}\right) \; ,
\end{align}
with the mass-squared matrices $M_{\pm}^{2}$, $M_{\rm even}^{2}$ and $M_{\rm odd}^{2}$ given by
\begin{eqnarray}
\label{10}
M_{\pm}^{2} \!=\! \left(\begin{matrix} \sqrt{2}\mu v_{\Delta} & -\mu v_{H} \cr -\mu v_{H} & M_{\Delta}^{2} \end{matrix}\right)  \; , 
M_{\rm even}^{2} \!=\! \left(\begin{matrix} \lambda v_{H}^{2}/2 & -\sqrt{2}\mu v_{H} \cr -\sqrt{2}\mu v_{H} & M_{\Delta}^{2} \end{matrix}\right)  \; , 
M_{\rm odd}^{2} \!=\! \left(\begin{matrix} 2\sqrt{2}\mu v_{\Delta} & -\sqrt{2}\mu v_{H} \cr -\sqrt{2}\mu v_{H} & M_{\Delta}^{2} \end{matrix}\right)  ,
\end{eqnarray}
which all are real and symmetric and can be diagonalized by orthogonal transformations.
To diagonalize the above mass-squared matrices,
we introduce three orthogonal matrices to rotate the Lagrangian fields into their mass eigenstates in the following way:
\begin{eqnarray}
\label{11}
\left(\begin{matrix} \phi^{\pm} \cr  \delta^{\pm} \end{matrix}\right) &=&
\left(\begin{matrix} \cos{\theta_{\pm}} & -\sin{\theta_{\pm}} \cr \sin{\theta_{\pm}} & \cos{\theta_{\pm}} \end{matrix}\right)
\left(\begin{matrix} G^{\pm} \cr  H^{\pm} \end{matrix}\right) \; , \quad {\rm with} \quad
\tan{\theta_{\pm}}=\frac{\sqrt{2}v_{\Delta}}{v_{H}}  \; , \nonumber \\
\left(\begin{matrix} \xi \cr  \zeta \end{matrix}\right) &=&
\left(\begin{matrix} \cos{\alpha} & -\sin{\alpha} \cr \sin{\alpha} & \cos{\alpha} \end{matrix}\right)
\left(\begin{matrix} h^{0} \cr  H^{0} \end{matrix}\right) \; , \quad {\rm with} \quad
\tan{2\alpha}=\frac{4v_{\Delta}/v_{H}}{1-\lambda v_{\Delta}/\left(\sqrt{2}\mu\right)}  \; , \nonumber \\
\left(\begin{matrix} \chi \cr  \eta \end{matrix}\right) &=&
\left(\begin{matrix} \cos{\beta} & -\sin{\beta} \cr \sin{\beta} & \cos{\beta} \end{matrix}\right)
\left(\begin{matrix} G^{0} \cr  A^{0} \end{matrix}\right) \; , \quad {\rm with} \quad
\tan{\beta}=\frac{2v_{\Delta}}{v_{H}}  \; .
\end{eqnarray}
After diagonalization, the Higgs mass spectrum can be given by
\begin{align}
\label{12}
&M_{H^{\pm\pm}}^{2} = M_{\Delta}^{2} \; , \quad M_{H^{\pm}}^{2} = M_{\Delta}^{2}\left(1+\frac{2v_{\Delta}^{2}}{v_{H}^{2}}\right) \; , \nn \\
&M_{h^{0}}^{2} = M_{\Delta}^{2}\left(\frac{\lambda v_{\Delta}}{\sqrt{2}\mu}\cos^{2}{\alpha}+\sin^{2}{\alpha}-\frac{2v_{\Delta}}{v_{H}}\sin{2\alpha}\right) \; , \nn \\
&M_{H^{0}}^{2} = M_{\Delta}^{2}\left(\frac{\lambda v_{\Delta}}{\sqrt{2}\mu}\sin^{2}{\alpha}+\cos^{2}{\alpha}+\frac{2v_{\Delta}}{v_{H}}\sin{2\alpha}\right) \; , \nn \\
&M_{A^{0}}^{2} = M_{\Delta}^{2}\left(1+\frac{4v_{\Delta}^{2}}{v_{H}^{2}}\right) \; , \quad M_{G^{\pm}}^{2} = M_{G^{0}}^{2} = 0 \; ,
\end{align}
where the doubly charged mass eigenstates $\delta^{\pm\pm}$ are replaced by $H^{\pm\pm}$. $G^{\pm}$ and $G^{0}$ correspond to the charged and neutral massless Goldstone bosons, which give masses to the electroweak gauge bosons $W^{\pm}$ and $Z$.
Taking the limit of $v_{\Delta}\ll v_{H}$, one can get a quasi-degenerate mass spectrum for the $\Delta$-like Higgs states
\begin{eqnarray}
\label{13}
M_{H^{\pm}}^{2} \simeq M_{H^{0}}^{2} \simeq M_{A^{0}}^{2} \simeq M_{H^{\pm\pm}}^{2} = M_{\Delta}^{2} \; .
\end{eqnarray}
In this minimal setting, the cascade decays between two heavy triplet Higgs bosons, such as
\begin{eqnarray}
\label{14}
H^{\pm\pm} \rightarrow H^{\pm}H^{\pm} \; ,  \quad
H^{\pm\pm} \rightarrow H^{\pm}W^{\pm} \; ,  \quad
H^{\pm} \rightarrow H^{0}W^{\pm}/A^{0}W^{\pm} \; ,  \quad
H^{0} \rightarrow A^{0}Z \; ,
\end{eqnarray}
are kinematically forbidden.

Furthermore, the tiny neutrino masses can be generated from the Yukawa interaction between the lepton doublet $\ell^{}_{\rm L}$ and the Higgs triplet $\Delta$
\begin{eqnarray}
\label{15}
{\cal L}_{\rm Yukawa}=-Y_{\nu} \overline{(\ell_{L})^{c}} i \sigma^{2} \Delta \ell_{L}+{\rm h.c.} \; ,
\end{eqnarray}
where $\overline{(\ell_{L})^{c}}=(\ell_{L})^{T}C$ with $C$ being the charge-conjugation operator, and $Y_\nu$ is the $3\times3$ neutrino Yukawa coupling matrix.
After spontaneous gauge symmetry breaking, the effective Majorana neutrino mass matrix can be given by
\begin{eqnarray}
\label{16}
M_{\nu}=\sqrt{2}Y_{\nu}v_{\Delta} = Y_{\nu}\frac{\mu v_{H}^{2}}{M_{\Delta}^{2}} \; .
\end{eqnarray}
Here the cut-off scale $\Lambda$ of new physics is replaced by $M_{\Delta}^{2}/\mu$.
If $\mu\ll M_{\Delta}$, the smallness of the neutrino masses can be explained by the seesaw spirit.
In this paper, in order to search for the existence of the introduced Higgs triplet at future $e^{-}p$ colliders, we explore the production and decay of the triplet scalars from the phenomenological point of view.
In the basis where the mass eigenstates of the charged leptons are identified with their flavor eigenstates, the effective neutrino mass matrix can be diagonalized by the so-called Pontecorvo-Maki-Nakagawa-Sakata (PMNS) matrix~\cite{Pontecorvo:1957cp,Maki:1962mu}, for which the standard parametrization~\cite{Tanabashi:2018oca} can be given by
\begin{eqnarray}
\label{17}
V_{\rm PMNS} \!=\! \left( \begin{array}{ccc}
c^{}_{13} c^{}_{12} & c^{}_{13} s^{}_{12} & s^{}_{13} e^{-{i}\delta} \\
-s_{12}^{} c_{23}^{} - c_{12}^{} s_{13}^{} s_{23}^{} e^{{i}\delta}_{} & + c_{12}^{} c_{23}^{} - s_{12}^{} s_{13}^{} s_{23}^{} e^{{i}\delta}_{} & c_{13}^{} s_{23}^{} \\
+ s_{12}^{} s_{23}^{} - c_{12}^{} s_{13}^{} c_{23}^{} e^{{i}\delta}_{} & - c_{12}^{} s_{23}^{} - s_{12}^{} s_{13}^{} c_{23}^{} e^{{i}\delta}_{} & c_{13}^{} c_{23}^{} \end{array} \right)
\!\times\! {\rm diag}\left(1,e^{{i}\alpha_{1}/2},e^{{i}\alpha_{2}/2}\right) ,
\end{eqnarray}
where $c^{}_{ij} \equiv \cos \theta^{}_{ij}$ and $s^{}_{ij} \equiv \sin \theta^{}_{ij}$ have been defined. $\delta$ is the Dirac CP-violating phase, while $\alpha_{1}$ and $\alpha_{2}$ are two Majorana CP-violating phases.
Using Eq.~(\ref{16}) one can rewrite the neutrino Yukawa coupling matrix as
\begin{eqnarray}
\label{18}
Y_{\nu}=\frac{M_{\nu}}{\sqrt{2}v_{\Delta}} = \frac{V_{\rm PMNS}^{\ast}\widehat{M_{\nu}}V_{\rm PMNS}^{\dagger}}{\sqrt{2}v_{\Delta}} \; ,
\end{eqnarray}
where $\widehat{M_{\nu}}={\rm Diag}\{m_{1}, m_{2}, m_{3}\}$ with $m_i$ ($i=1,2,3$) being the neutrino mass eigenvalues.
The values of $Y_{\nu}$ are thus governed by the neutrino oscillation parameters and the Higgs triplet vev.
It is also worth mentioning that due to the simultaneous existence of the $\mu$ term in Eq.~(\ref{4}) and the Yukawa interaction term in Eq.~(\ref{15}), one finds that the lepton number in this model is explicitly violated by two units, so we can probe the production signal of the introduced Higgs triplet via the lepton-number violating processes.

\section{Constraints on the model parameters}\label{sec3}

\subsection{Constraints from neutrino oscillation experiments}\label{subsec3.1}

The latest global analysis of neutrino oscillation data~\cite{Esteban:2018azc} yields the best-fit values of the neutrino oscillation parameters
\begin{align}
\label{19}
\sin^2 \theta^{}_{12} \simeq 0.310  \; , \quad  \sin^2 \theta^{}_{23} \simeq 0.563  \; , \quad & \sin^2 \theta^{}_{13} \simeq 0.02237 \; ,  \nn \\
\Delta m^2_{21} \simeq 7.39\times 10^{-5}~{\rm eV}^2  \; , \quad \Delta m^2_{31} \simeq 2.528 & \times 10^{-3}~{\rm eV}^2  \; , \quad \delta = 221^\circ \; ,
\end{align}
where only the normal ordering of neutrino masses (i.e., $m^{}_1 < m^{}_2 < m^{}_3$) is considered for illustration purpose.
The absolute scale of neutrino masses has not yet been determined experimentally, and the updated upper limit on the sum of the neutrino masses reported by the Planck collaboration is $m_{1} + m_{2} + m_{3} < 0.12~{\rm eV}$~\cite{Aghanim:2018eyx}.
For simplicity, the mass of the lightest neutrino is set to zero (i.e., $m^{}_1=0$) in our numerical analysis, which is consistent with the experimental constraint~\cite{Esteban:2018azc,Aghanim:2018eyx}.
Since the neutrino oscillation probabilities are independent of the Majorana CP-violating phases, we further neglect the effects of the Majorana phases (i.e., $\alpha_{1}=\alpha_{2}=0$).

\subsection{Constraints from lepton flavor violating processes}\label{subsec3.2}

The charged Higgs bosons introduced in this model may induce many rare lepton flavor violating decays such as $\ell_{\alpha}\to\ell_{\beta}\ell_{\gamma}\ell_{\delta}$ and $\ell_{\alpha}\to\ell_{\beta}\gamma$~\cite{Pal:1983bf,Leontaris:1985qc,Swartz:1989qz,Mohapatra:1992uu,Akeroyd:2009nu}.
The experimental limits on the branching ratios of these various lepton flavor violating processes can be used to set some stringent constraints on the neutrino Yukawa coupling, and the most stringent bounds can be derived from the $\mu \rightarrow e\gamma$ (mediated by $H^{\pm\pm}$ and $H^{\pm}$) and $\mu \rightarrow 3e$ (mediated by $H^{\pm\pm}$) decays:
\begin{itemize}
  \item the branching ratio for $\mu \rightarrow e\gamma$ can be expressed as
  \begin{eqnarray}
  \label{20}
  {\rm BR}\left(\mu \rightarrow e\gamma\right)\simeq\frac{27\alpha|\left(Y_{\nu}^{\dagger}Y_{\nu}\right)^{e\mu}|^2}{64\pi G^{2}_{F}M_{\Delta}^{4}} \; ,
  \end{eqnarray}
  where $\alpha$ is the fine structure constant and $G_{F}$ is the Fermi coupling constant. The current experimental limit ${\rm BR}\left(\mu \rightarrow e\gamma\right) < 4.2 \times 10^{-13}$ ($90\%$ C.L.)~\cite{TheMEG:2016wtm} requires that
  \begin{eqnarray}
  \label{21}
  |\left(Y_{\nu}^{\dagger}Y_{\nu}\right)^{e\mu}|< 2.4 \times 10^{-6}\times\left(\frac{M_{\Delta}}{100~{\rm GeV}}\right)^2 \; ,
  \end{eqnarray}
  \item the branching ratio for $\mu \rightarrow 3e$ is given by
  \begin{eqnarray}
  \label{22}
  {\rm BR}\left(\mu \rightarrow 3e\right)\simeq\frac{|Y_{\nu}^{\mu e}|^2 |Y_{\nu}^{ee}|^2}{4\pi G^{2}_{F}M_{\Delta}^{4}} \; .
  \end{eqnarray}
  Making use of the current experimental bound ${\rm BR}\left(\mu \rightarrow 3e\right) < 1.0 \times 10^{-12}$ ($90\%$ C.L.)~\cite{Bellgardt:1987du}, one can easily find that
  \begin{eqnarray}
  \label{23}
  |Y_{\nu}^{\mu e}| |Y_{\nu}^{ee}| < 2.3 \times 10^{-7}\times\left(\frac{M_{\Delta}}{100~{\rm GeV}}\right)^2 \; .
  \end{eqnarray}
\end{itemize}
In addition, the doubly charged Higgs boson also contributes to the anomalous magnetic moment of electron and muon, muonium-antimuonium conversion, and $e^{}e^{}\rightarrow \ell^{}\ell^{}$ scattering, which give much weaker limits on the Yukawa coupling, see Refs.~\cite{Cuypers:1996ia,Dev:2018kpa} for reviews.

\subsection{Constraints from electroweak precision measurements}\label{subsec3.3}

As mentioned above, the Higgs triplet vev can contribute to the electroweak gauge boson masses at tree-level through
\begin{eqnarray}
\label{24}
M_{W}^{2}=\frac{g^{2}}{4}\left(v_{H}^{2}+2v_{\Delta}^{2}\right) \; , \quad
 M_{Z}^{2}=\frac{g^{2}}{4\cos^{2}{\theta_{W}}}\left(v_{H}^{2}+4v_{\Delta}^{2}\right) \; ,
\end{eqnarray}
with $\theta_{W}$ being the Weinberg angle. Thus the $\rho$ parameter in this model can be expressed as
\begin{eqnarray}
\label{25}
\rho=\frac{M_{W}^{2}}{M_{Z}^{2}\cos^{2}{\theta_{W}}}= \frac{1+2v_{\Delta}^{2}/v_{H}^{2}}{1+4v_{\Delta}^{2}/v_{H}^{2}}.
\end{eqnarray}
The electroweak precision measurement of the $\rho$ parameter requires that $v_{\Delta}/v_{H}\lesssim 0.03$ or $v_{\Delta}<8~{\rm GeV}$~\cite{Akeroyd:2007zv}.
As shown in previous subsection, the neutrino Yukawa coupling is strongly constrained by the lepton flavor violating processes, which in turn gives a lower bound on the Higgs triplet vev with the help of Eq.~(\ref{16}). In the conservative case, the value of the Higgs triplet vev used in our numerical analysis is assumed to be
\begin{eqnarray}
\label{26}
10~{\rm eV} \lesssim v_{\Delta} \lesssim 1~{\rm GeV} \; .
\end{eqnarray}

\subsection{Constraints from the LHC experiments}\label{subsec3.4}

Direct searches for the triplet scalars have been carried out at LHC for various production and decay modes.
No significant deviations from the standard model predictions are found and lower limits on the triplet masses are derived at 95\% confidence level.
At present, the most stringent constraints are mainly from the searches for the doubly charged Higgs.
The ATLAS collaboration has recently searched for $H^{\pm\pm}$ via the Drell-Yan process and released its preliminary results with an integrated luminosity of 36.1 ${\rm fb}^{-1}$ collected at $\sqrt{s}=13~{\rm TeV}$. For the $\ell^{\pm}\ell^{\pm}~(e^{\pm}e^{\pm}/\mu^{\pm}\mu^{\pm}/e^{\pm}\mu^{\pm})$ channel, the observed lower limit on the mass of $H^{\pm\pm}$ varies from 770 GeV to 870 GeV for ${\rm BR}(H^{\pm\pm} \rightarrow\ell^{\pm}\ell^{\pm}) = 100\%$ and is above 450 GeV for ${\rm BR}(H^{\pm\pm} \rightarrow\ell^{\pm}\ell^{\pm}) \geq 10\%$~\cite{Aaboud:2017qph}.
For the $W^{\pm}W^{\pm}$ channel, the observed lower limit on the mass of $H^{\pm\pm}$ is 220 GeV~\cite{Aaboud:2018qcu}.
The CMS collaboration has also searched for $H^{\pm\pm}$ in the pair production mode $pp\rightarrow H^{++}H^{--}\rightarrow\ell^{+}_{}\ell^{+}_{}\ell^{-}_{}\ell^{-}_{}$ and the associated production mode $pp\rightarrow H^{\pm\pm}H^{\mp}\rightarrow\ell^{\pm}_{}\ell^{\pm}_{}\ell^{\mp}_{}\nu_{}$ with an integrated luminosity of 12.9 ${\rm fb}^{-1}$ collected at $\sqrt{s}=13~{\rm TeV}$.
The lower bounds on the $H^{\pm\pm}$ mass are established between 535 GeV and 820 GeV in the 100\% branching ratio scenarios, and between 716 GeV and 761 GeV for four benchmark points of the type-II seesaw model~\cite{CMS:2017pet}.
For singly charged $H^{\pm}$ and neutral $H^{0}/A^{0}$, the summary of current constraints from the LHC experiments can be found in Ref.~\cite{Flechl:2019dtr}.

\section{Production and decay of the triplet scalars}\label{sec4}

In this section, we first discuss the dominant production channels of the triplet scalars via vector boson fusion process at $e^{-}p$ colliders and subsequently study their decay properties.

\subsection{Production of the triplet scalars}\label{subsec4.1}

\begin{figure}[!htbp]
\begin{center}
\subfigure[]{\label{fig1a}
\includegraphics[width=0.34\textwidth]{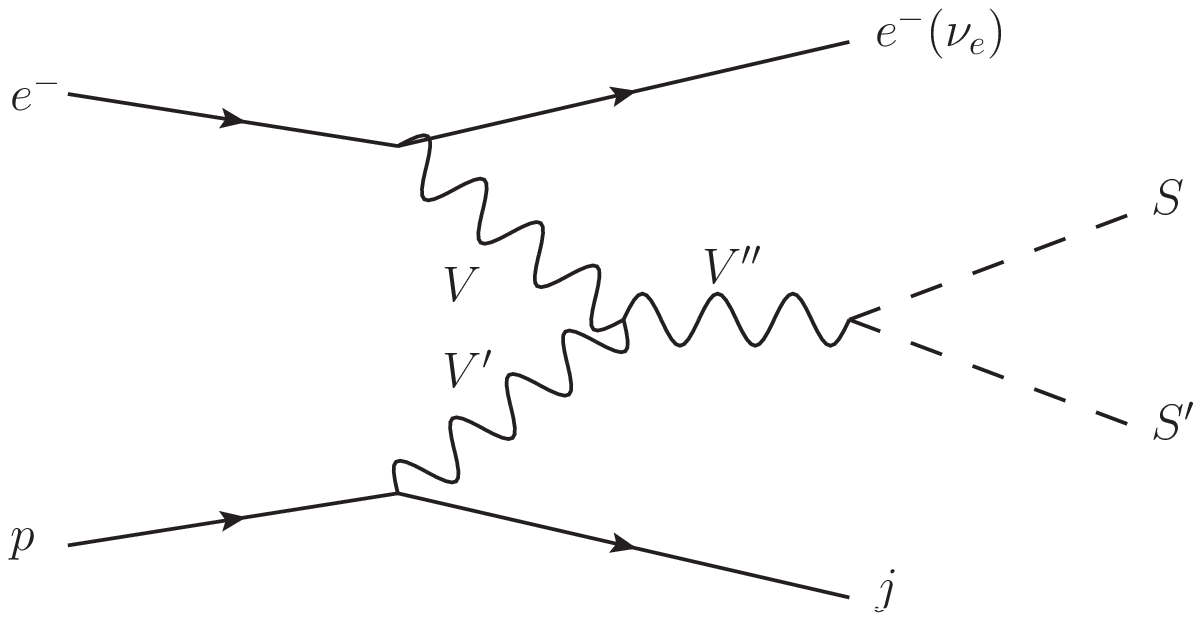} }
\hspace{-0.5cm}~
\subfigure[]{\label{fig1b}
\includegraphics[width=0.30\textwidth]{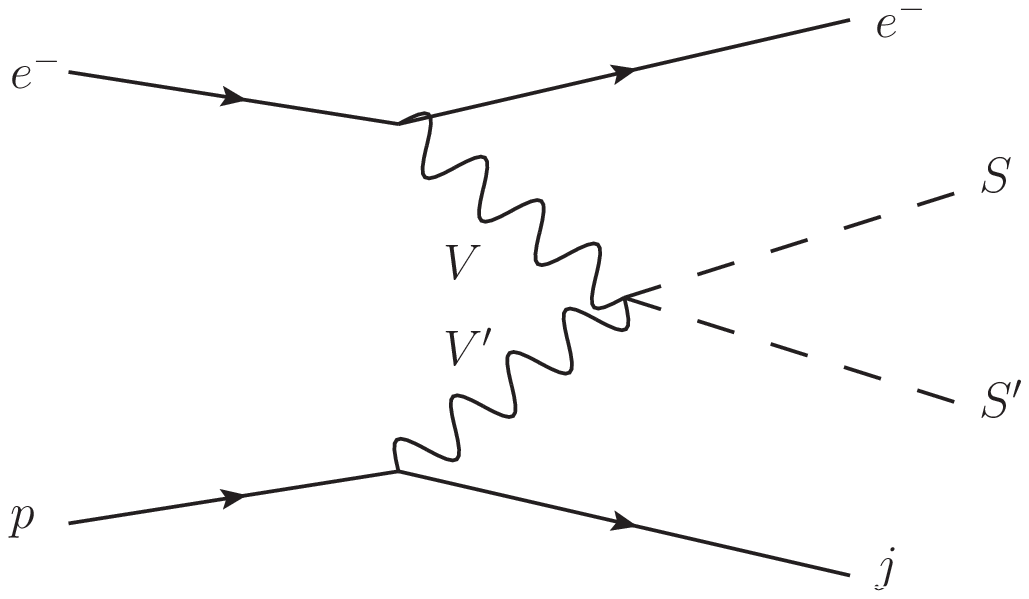} }
\hspace{-0.5cm}~
\subfigure[]{\label{fig1c}
\includegraphics[width=0.28\textwidth]{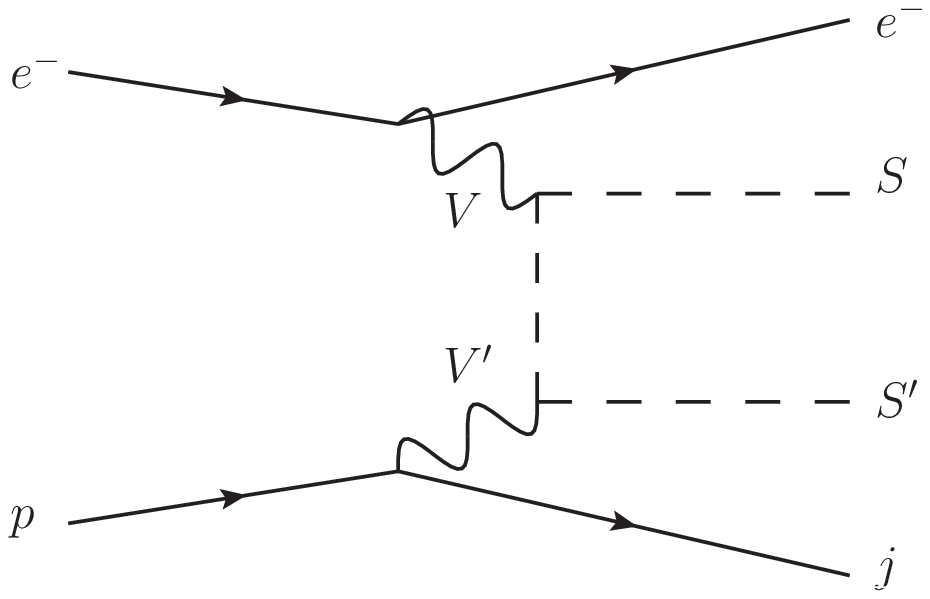} }
\caption{Feynman diagrams for the production of a pair of triplet scalars via vector boson fusion process at $e^{-}p$ colliders, where the vector bosons $V$, $V^{\prime}$ and $V^{\prime\prime}$ can be $W$, $Z$ or $\gamma$.}\label{fig1}
\end{center}
\end{figure}

For vector boson fusion process, the triplet scalars can be produced by the fusion of two virtual vector bosons such as $W$, $Z$ or $\gamma$.
Since the single production of the triplet scalars is highly suppressed by the small Higgs triplet vev, we only consider here the production of a pair of triplet scalars through
\begin{eqnarray}
\label{27}
e^{-} + p \rightarrow e^{-} + S + S^{\prime} + j \; ,
\end{eqnarray}
and
\begin{eqnarray}
\label{28}
e^{-} + p \rightarrow \nu_{e} + S + S^{\prime} + j \; ,
\end{eqnarray}
with $S$, $S^{\prime}$ being $H^{\pm\pm}$, $H^{\pm}$, $H^{0}$ or $A^{0}$ (see Fig.~\ref{fig1}).
As the photon-mediated processes receive contributions from the collinear photon, they should be treated with great care.
Specifically, for the photon emitted from the electron, we employ the following photon density function~\cite{Frixione:1993yw}
\begin{eqnarray}
\label{29}
f_{\gamma/e^-}(x)=
\frac{\alpha}{2\pi}\left[\frac{1+(1-x)^2}{x}\ln{\frac{Q_{\rm max}^2}{Q_{\rm min}^2}} + 2 m_e^2 x \left(\frac{1}{Q_{\rm max}^2}-\frac{1}{Q_{\rm min}^2}\right)\right] \; ,
\end{eqnarray}
where $Q_{\rm min}^2=m_{e}^2 x^2/(1-x)$, $Q_{\rm max}^2=(\theta_{c} E_{e})^2(1-x)+Q_{\rm min}^2$ with $x$ the energy fraction of the photon and $E_{e}$ the energy of the electron, $m_e=0.51~{\rm MeV}$ the mass of electron and $\theta_c=32~{\rm mrad}$ the cut of the electron scattering angle.
As for the photon emitted from the proton, the photon parton distribution function $f_{\gamma/p}(x, \mu^2_{f})$ that includes both elastic and inelastic contributions is adopted, where $\mu_{f}$ characterizes the factorization scale.
Here, the CT14QED~\cite{Schmidt:2015zda} parton distribution functions are employed in the calculation, and the factorization scale is set at $\sqrt{\hat{s}}$ with $\hat{s}$ the partonic center-of-mass energy.

\begin{figure}[!htbp]
\begin{center}
\subfigure[]{\label{fig2a}
\includegraphics[width=0.40\textwidth]{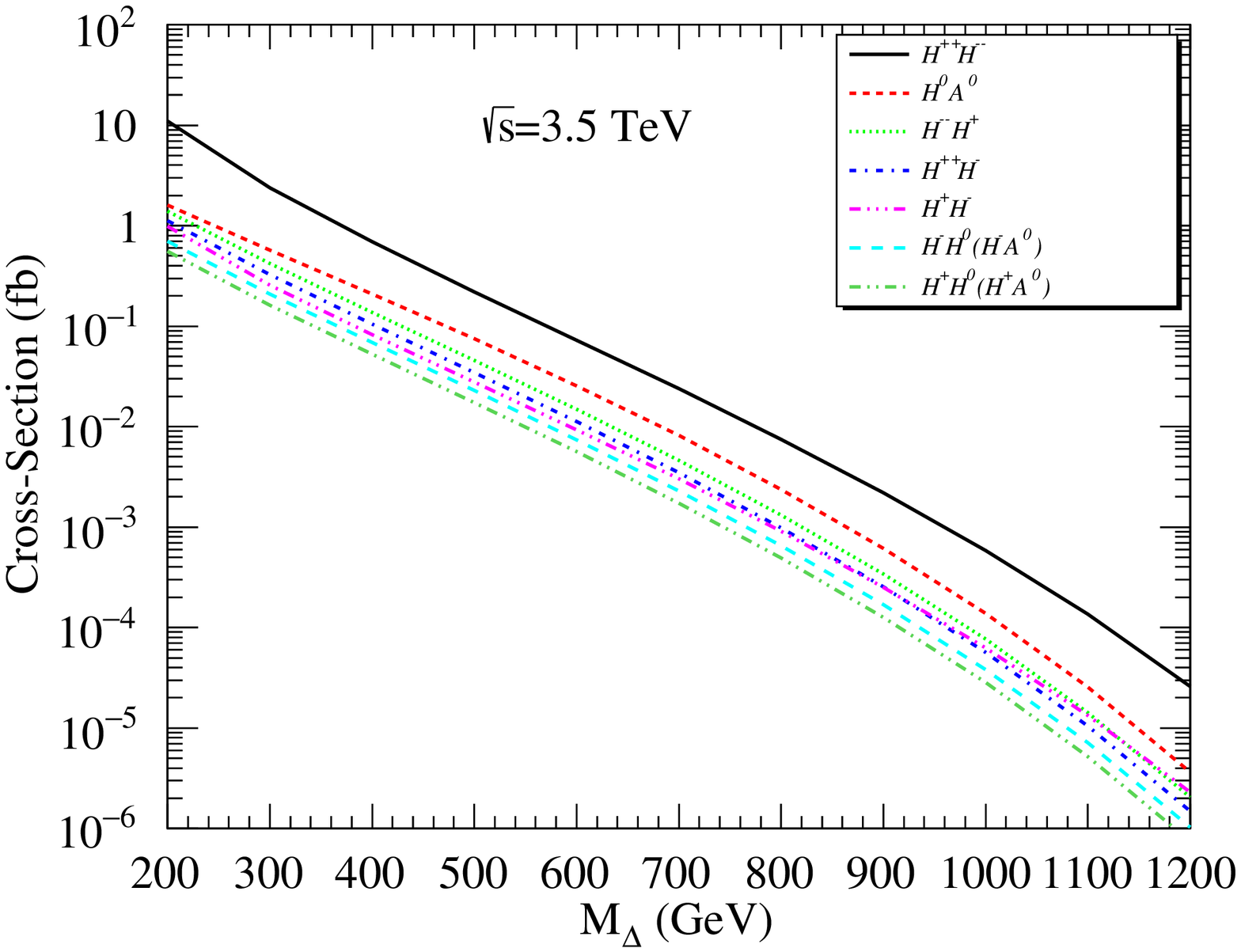} }
\hspace{-0.5cm}~
\subfigure[]{\label{fig2b}
\includegraphics[width=0.40\textwidth]{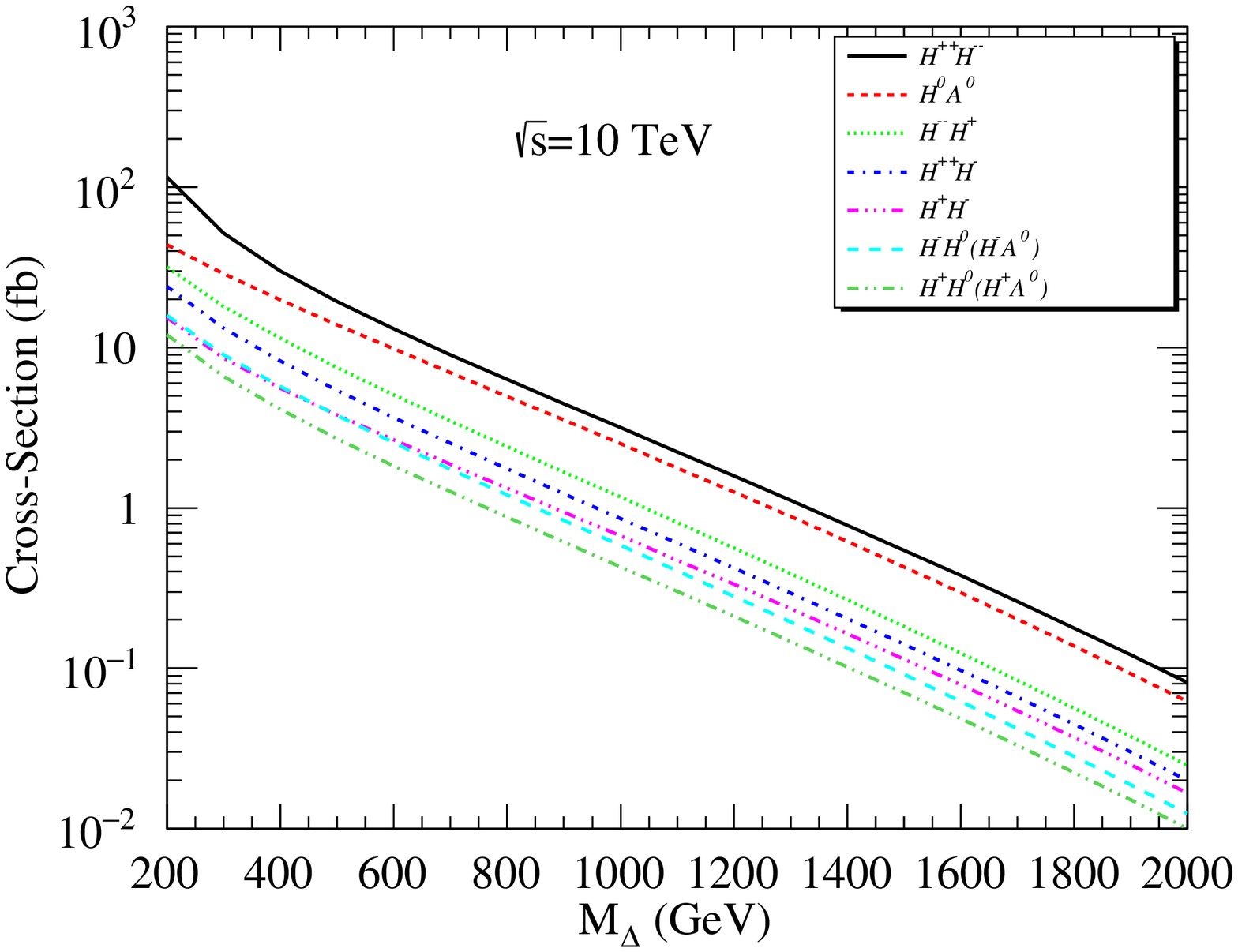} }
\caption{The inclusive production cross sections of a pair of triplet scalars via vector boson fusion process at (a) FCC-ep and (b) ${\rm ILC}\otimes{\rm FCC}$ as a function of $M_{\Delta}$.}\label{fig2}
\end{center}
\end{figure}

Before giving the simulated results, we briefly summarize our simulation procedures.
For the signal processes, we develop a package by the help of Form~\cite{Kuipers:2012rf} to generate a Fortran code, and take advantage of Vegas~\cite{Lepage:1980dq} to perform the numerical integration. While the backgrounds in the standard model are simulated by MadGraph~\cite{Alwall:2014hca}.

The total production cross sections of a pair of triplet scalars for the processes in Eq.~(\ref{27}) and Eq.~(\ref{28}) at FCC-ep with $\sqrt{s}=3.5~{\rm TeV}$ and ${\rm ILC}\otimes{\rm FCC}$ with $\sqrt{s}=10~{\rm TeV}$ are shown in Fig.~\ref{fig2} as a function of the triplet scalar mass.
It is easy to find that the pair production of doubly charged Higgs ($H^{++}H^{--}$) has the largest production cross sections and will be used as the discovery channel to investigate the collider phenomenology of the Higgs triplet at future $e^{-}p$ colliders in next section.

\subsection{Decay of the triplet scalars}\label{subsec4.2}

\begin{figure}[!htbp]
\begin{center}
\subfigure[]{\label{fig3a}
\includegraphics[width=0.40\textwidth]{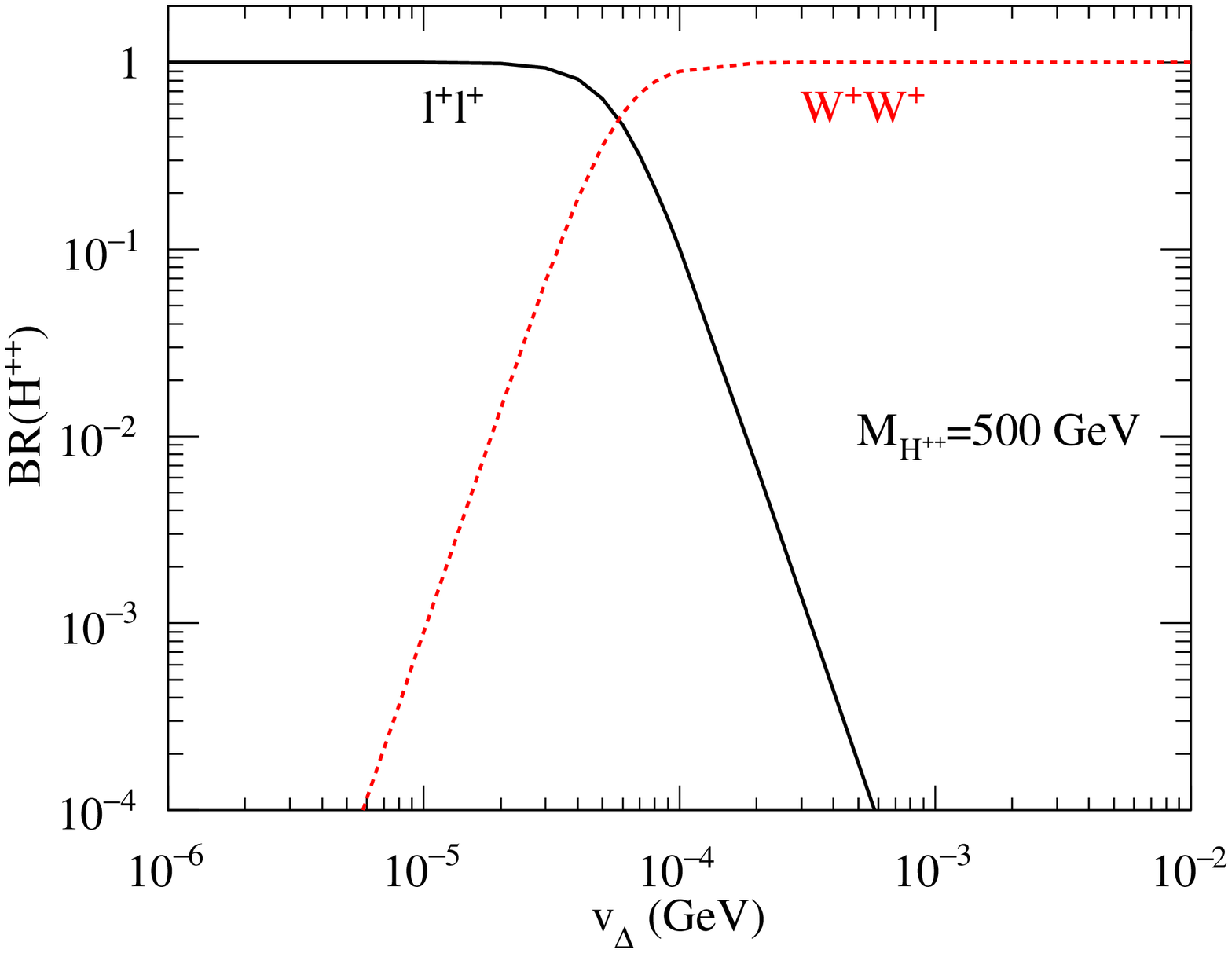} }
\hspace{-0.5cm}~
\subfigure[]{\label{fig3b}
\includegraphics[width=0.40\textwidth]{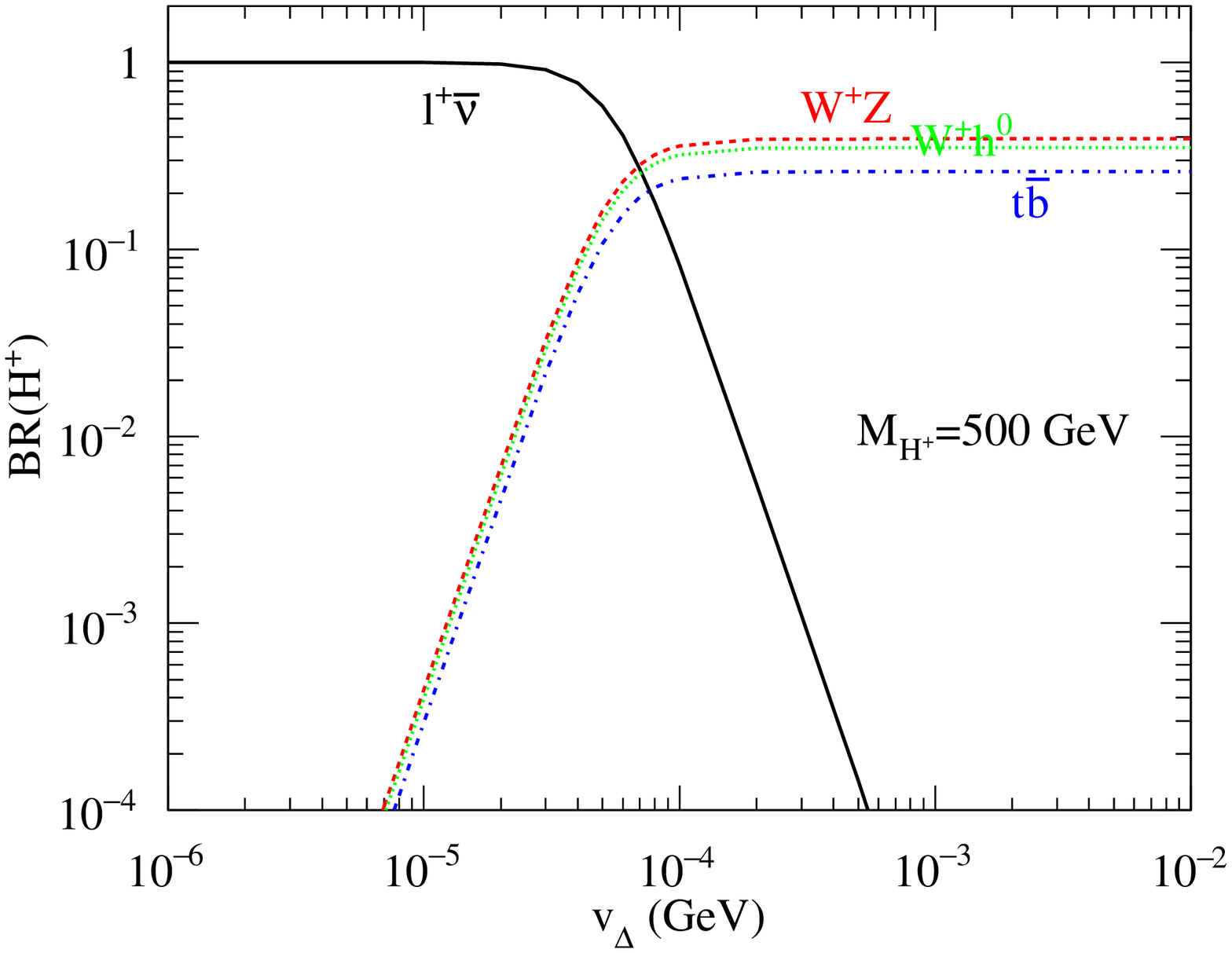} }
\hspace{-0.5cm}~
\subfigure[]{\label{fig3c}
\includegraphics[width=0.40\textwidth]{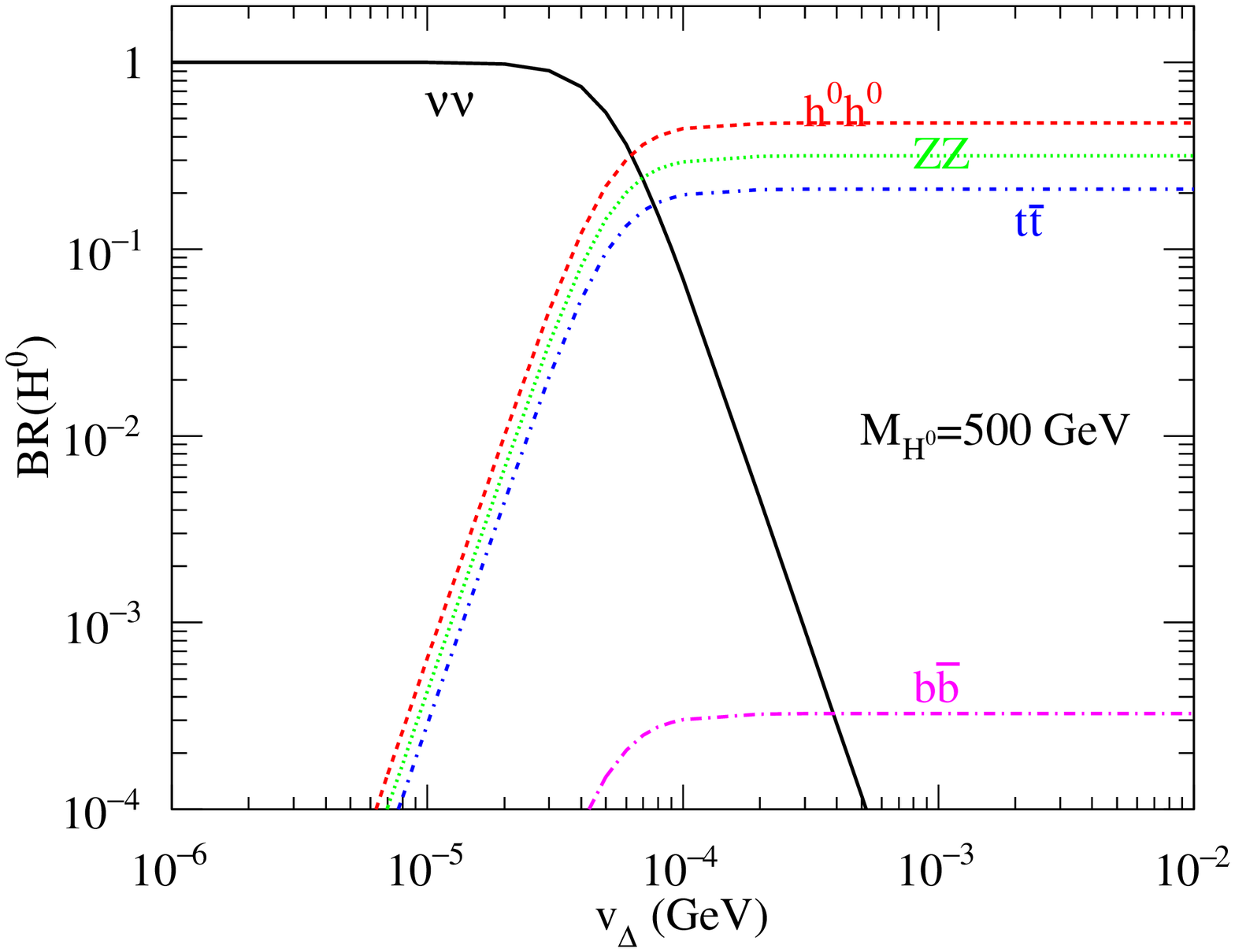} }
\hspace{-0.5cm}~
\subfigure[]{\label{fig3d}
\includegraphics[width=0.40\textwidth]{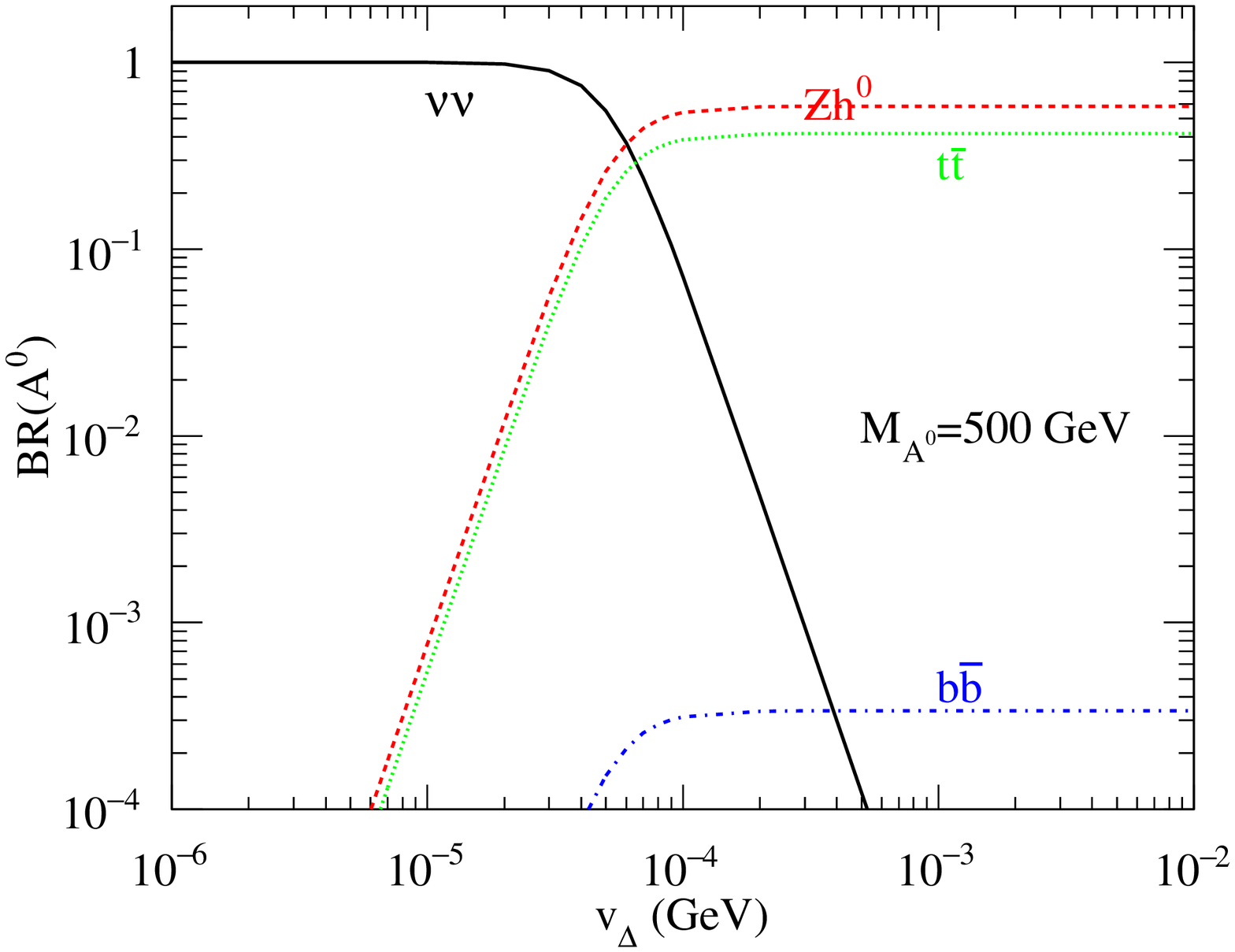} }
\caption{Branching ratios of (a) $H^{++}$, (b) $H^{+}$, (c) $H^{0}$ and (d) $A^{0}$ as a function of $v_{\Delta}$ for $M_{\Delta}=500~{\rm GeV}$.}\label{fig3}
\end{center}
\end{figure}

The decay rates of the triplet scalars are respectively sensitive to the Yukawa coupling and the Higgs triplet vev, which are connected by the relation in Eq.~(\ref{16}).
In Fig.~\ref{fig3}, we plot the branching ratios of $H^{++}$, $H^{+}$, $H^{0}$ and $A^{0}$ as a function of $v_{\Delta}$ for $M_{\Delta}=500~{\rm GeV}$ with the corresponding partial decay widths given in Ref.~\cite{Perez:2008ha}.
As shown in Fig.~\ref{fig3}, the possible decays of $H^{++}$ are $H^{++} \rightarrow \ell^{+}\ell^{+}$ and $H^{++} \rightarrow W^{+}W^{+}$.
For $M_{H^{++}}=500~{\rm GeV}$, the same-sign dilepton decay dominates for small values of $v_{\Delta}$, while the same-sign diboson decay becomes important for large values of $v_{\Delta}$.
In the case of $H^{+}$ with $M_{H^{+}}=500~{\rm GeV}$, the most relevant decay channel for small $v_{\Delta}$ is $H^{+} \rightarrow \ell^{+}\nu$, while $H^{+} \rightarrow W^{+}Z$, $W^{+}h^{0}$ and $H^{+} \rightarrow t\bar{b}$ are the dominant channels for large $v_{\Delta}$.
For $H^{0}$ and $A^{0}$ with $M_{H^{0},A^{0}}=500~{\rm GeV}$, the invisible decays $H^{0} \rightarrow \nu\nu$ and $A^{0} \rightarrow \nu\nu$ are the most important channels for small $v_{\Delta}$, while $H^{0} \rightarrow h^{0}h^{0}$, $ZZ$, $t\bar{t}$ and $A^{0} \rightarrow Zh^{0}$, $t\bar{t}$ become dominant for large $v_{\Delta}$.
All the decay properties of the triplet scalars will provide useful information for the collider study in next section.

\section{Discovery potential at future $ep$ colliders }\label{sec5}

As discussed in last section, the most important production channel of a pair of triplet scalars via vector boson fusion process at $e^{-}p$ colliders is the pair production of doubly charged Higgs, and the produced doubly charged Higgs may decay into a pair of same-sign charged leptons or a pair of same-sign $W$ bosons depending on the size of the Higgs triplet vev.
In order to explore the discovery potential of the Higgs triplet at future $e^{-}p$ colliders, in the following, we consider the inclusive production of the doubly charged Higgs through $e^{-}p\rightarrow H^{++} H^{--} + X$ with $H^{++}/H^{--}\rightarrow\ell^{+}\ell^{+}/\ell^{-}\ell^{-}$ and $H^{++}/H^{--}\rightarrow W^{+}W^{+}/W^{-}W^{-}$, respectively.

\subsection{Same-sign dilepton decay mode}\label{subsec5.1}

\begin{figure}[!htbp]
\begin{center}
\includegraphics[width=0.40\textwidth]{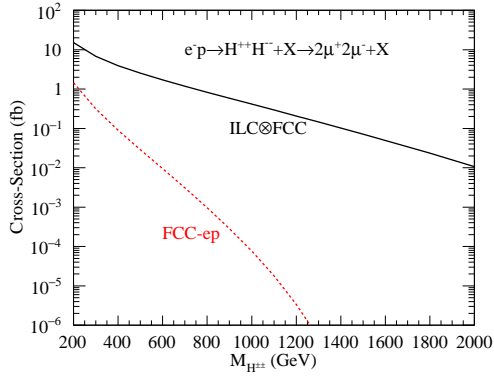}
\caption{ The inclusive production cross sections for $e^{-}p\rightarrow H^{++} H^{--} + X\rightarrow 2\mu^{+}2\mu^{-}+ X$ at FCC-ep and ${\rm ILC}\otimes{\rm FCC}$ as a function of $M_{H^{\pm\pm}}$.}\label{fig4}
\end{center}
\end{figure}

We first consider the case of the doubly charged Higgs decays dominantly into a pair of same-sign charged leptons:
\begin{eqnarray}
\label{30}
e^{-} + p &\rightarrow& e^{-} + H^{++} + H^{--} + j  \; , \quad {\rm with} \quad H^{++} \rightarrow \ell^{+}\ell^{+} \; , \; H^{--} \rightarrow  \ell^{-}\ell^{-} \; , \nn \\
e^{-} + p &\rightarrow& \nu_{e} + H^{++} + H^{--} + j  \; , \quad {\rm with} \quad H^{++} \rightarrow \ell^{+}\ell^{+} \; , \;H^{--} \rightarrow  \ell^{-}\ell^{-} \; .
\end{eqnarray}
The signal in this case consists of at least two same-sign dilepton pairs.
The decay branching ratios of the doubly charged Higgs to different flavors can be easily computed by using the best-fit values of the neutrino oscillation parameters in Eq.~(\ref{19}) with the assumption of ${\rm BR}(H^{\pm\pm}\rightarrow\ell^{\pm}\ell^{\pm})=100\%$ for small values of $v_{\Delta}$, and the numerical results are listed in Table~\ref{table1}.
For illustration, we only concentrate on the cleanest dimuon production mode.
Taking into account the decay branching ratios, we show the total cross sections for the inclusive process $e^{-}p\rightarrow H^{++} H^{--} + X\rightarrow 2\mu^{+}2\mu^{-}+ X$ at FCC-ep and ${\rm ILC}\otimes{\rm FCC}$ in Fig.~\ref{fig4} as a function of $M_{H^{\pm\pm}}$.

\begin{table}[!htbp]
  \caption{The decay branching ratios of doubly charged Higgs to different flavors.}\label{table1}
  \centering
  \begin{tabular}{p{1cm}<{\centering}p{1cm}<{\centering}p{1cm}<{\centering}p{1cm}<{\centering}p{1cm}<{\centering}p{1cm}<{\centering}p{1cm}<{\centering}}
  \hline
  \hline
   & $ee$ & $e\mu$ & $e\tau$ & $\mu\mu$ & $\mu\tau$ & $\tau\tau$ \\
  BR & 0.34\% & 1.28\% & 4.08\% & 36.2\% & 35.2\% & 22.9\%  \\
  \hline
  \hline
  \end{tabular}
\end{table}

To simulate the detector effects, we smear the lepton and jet energies according to the assumption of the Gaussian resolution parametrization
\begin{eqnarray}
\label{31}
\frac{\delta(E)}{E} = \frac{a}{\sqrt{E}}\oplus b,
\end{eqnarray}
where $\delta(E)/E$ is the energy resolution, $a$ is a sampling term, $b$ is a constant term, and $\oplus$ denotes a sum in quadrature. We take $a=5\%$, $b=0.55\%$ for leptons and $a=100\%$, $b=5\%$ for jets, respectively~\cite{Ball:2007zza,Aad:2009wy}.

\begin{figure}[!htbp]
\begin{center}
\subfigure[]{\label{fig5a}
\includegraphics[width=0.40\textwidth]{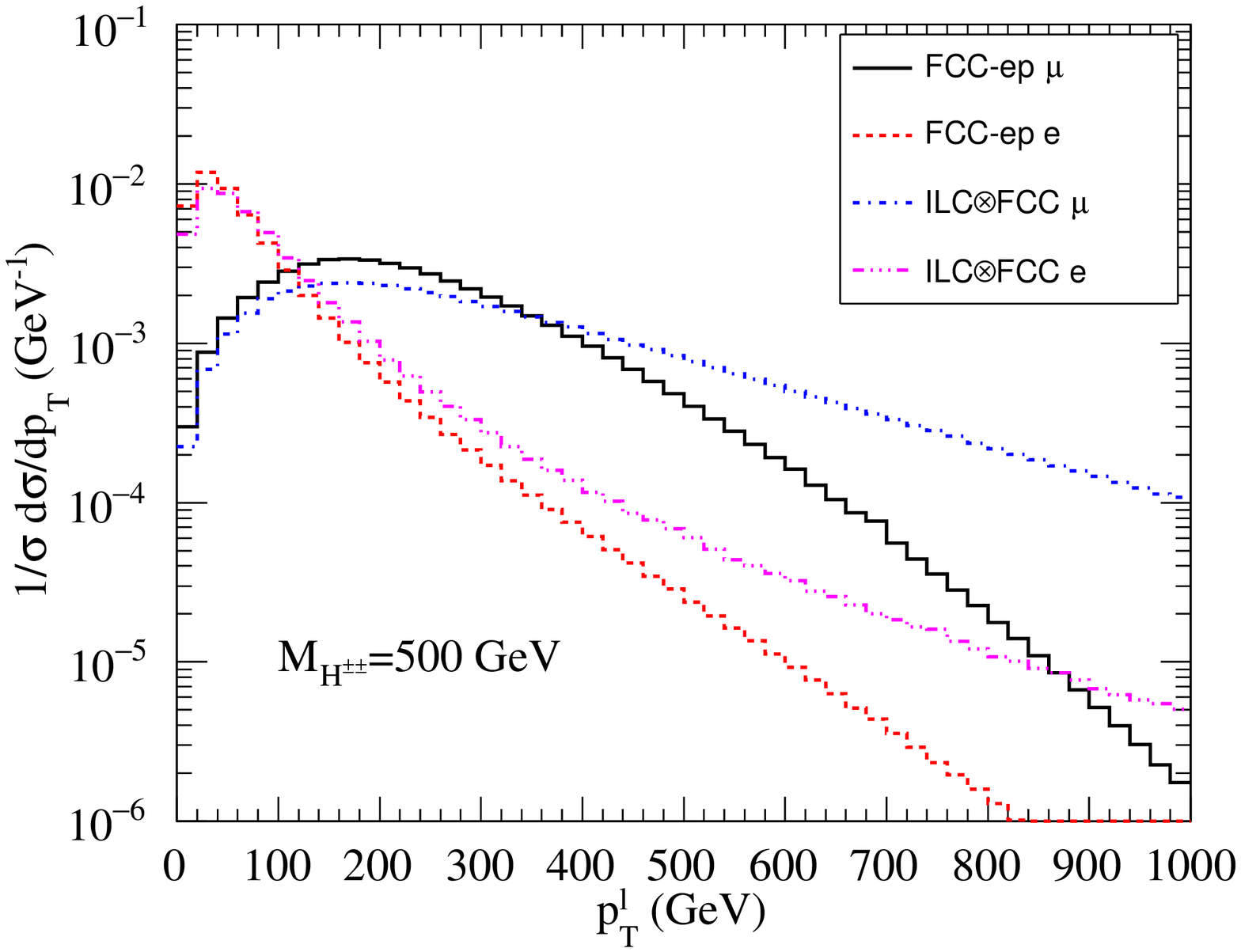} }
\hspace{-0.5cm}~
\subfigure[]{\label{fig5b}
\includegraphics[width=0.40\textwidth]{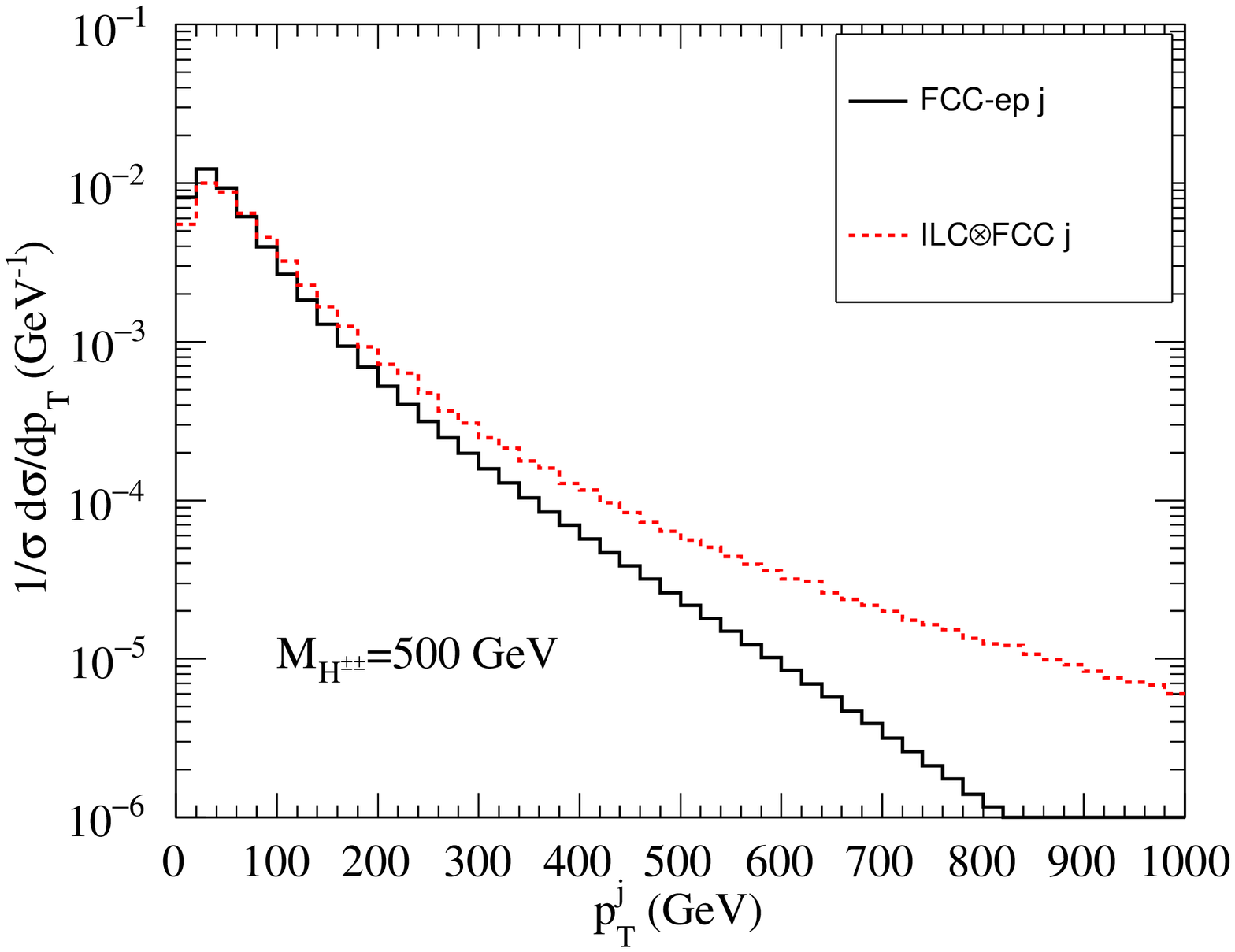} }
\caption{The normalized transverse momentum distributions (a) $1/\sigma{\rm d}\sigma/{\rm d}p^{\mu,e}_{\rm T}$ and (b) $1/\sigma{\rm d}\sigma/{\rm d}p^{j}_{\rm T}$ for $M_{H^{\pm\pm}}=500~{\rm GeV}$ at FCC-ep and ${\rm ILC}\otimes{\rm FCC}$.}\label{fig5}
\end{center}
\end{figure}

For the processes in Eq.~(\ref{30}), the four muons originating from $H^{\pm\pm}$ decay are labeled as $\mu_{i}$ ($i=1,2,3,4$) and are ranked by $p_{T}$ with $p_{T}^{\mu_{1}}>p_{T}^{\mu_{2}}>p_{T}^{\mu_{3}}>p_{T}^{\mu_{4}}$.
In order to investigate the transverse momentum distributions of the final state particles, it is useful to define the differential distribution of the four muons as $1/\sigma{\rm d}\sigma/{\rm d}p^{\mu}_{\rm T}=1/\sigma({\rm d}\sigma/{\rm d}p^{\mu_{1}}_{\rm T}+{\rm d}\sigma/{\rm d}p^{\mu_{2}}_{\rm T}+{\rm d}\sigma/{\rm d}p^{\mu_{3}}_{\rm T}+{\rm d}\sigma/{\rm d}p^{\mu_{4}}_{\rm T})/4$.
In Fig.~\ref{fig5}, we plot the normalized transverse momentum distributions $1/\sigma{\rm d}\sigma/{\rm d}p^{\mu,e,j}_{\rm T}$ for $M_{H^{\pm\pm}}=500~{\rm GeV}$ at FCC-ep and ${\rm ILC}\otimes{\rm FCC}$, respectively.

\begin{figure}[!htbp]
\begin{center}
\subfigure[]{\label{fig6a}
\includegraphics[width=0.40\textwidth]{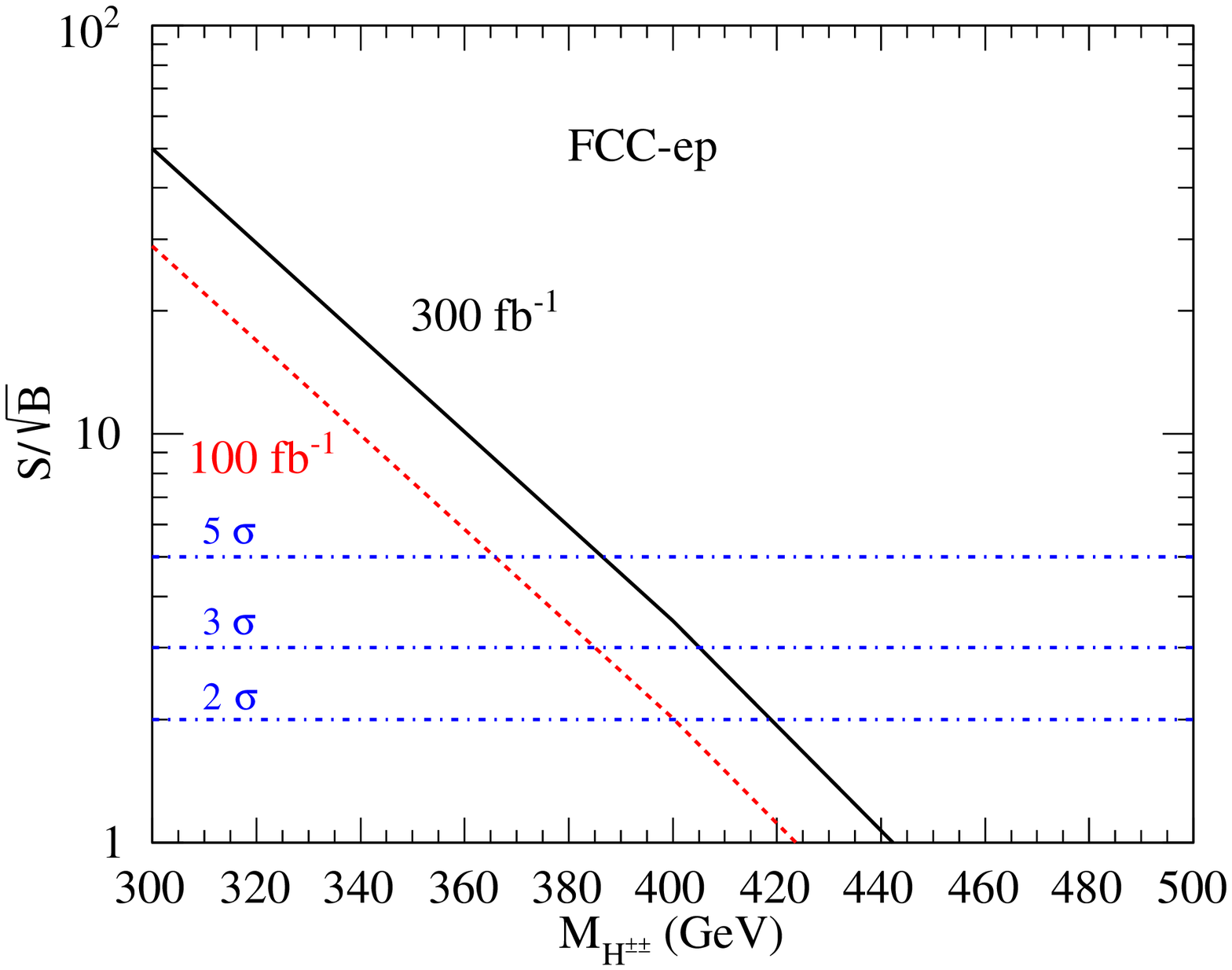} }
\hspace{-0.5cm}~
\subfigure[]{\label{fig6b}
\includegraphics[width=0.40\textwidth]{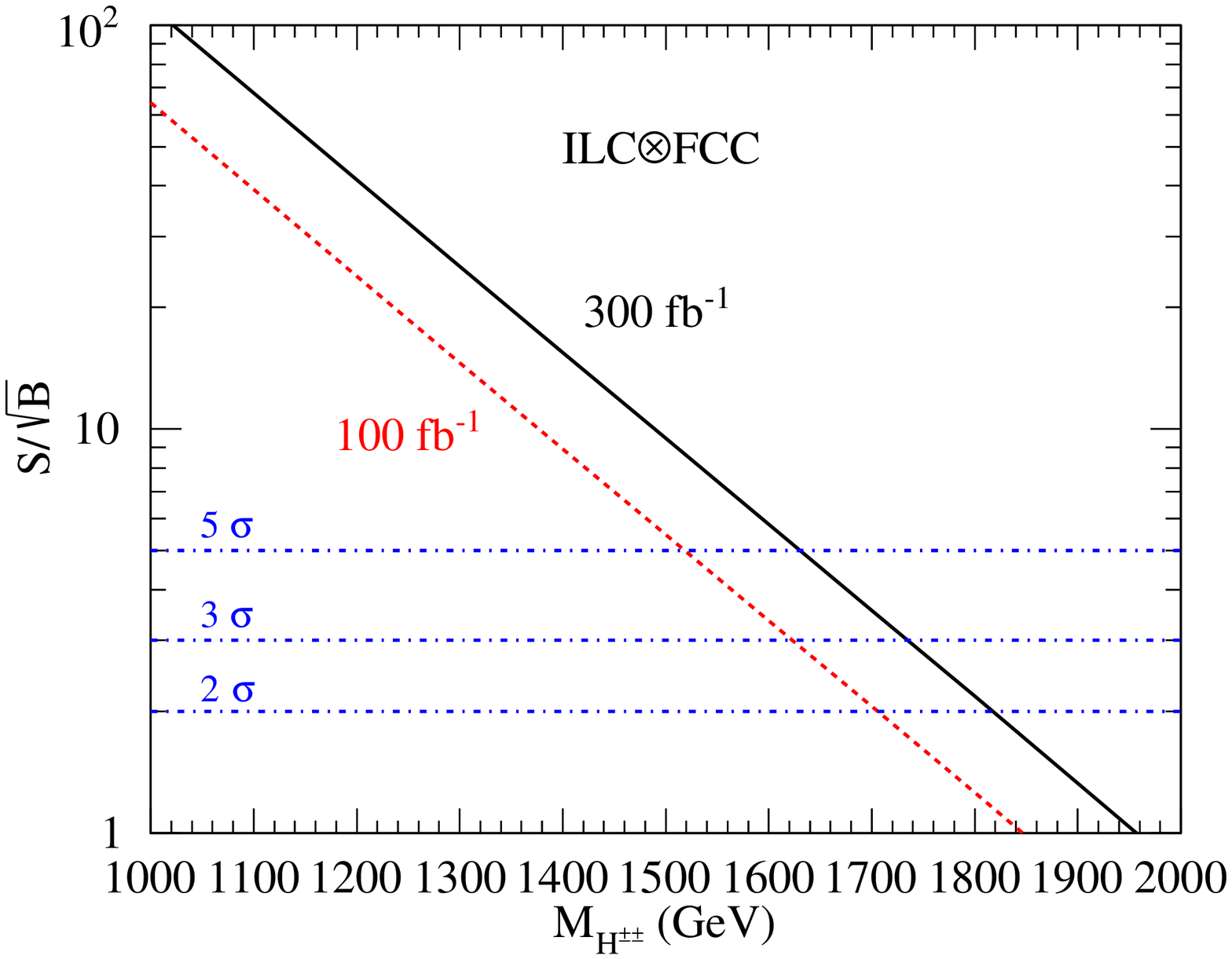} }
\caption{The statistical significance as a function of the doubly charged Higgs mass with the integraetd luminosity of $100~{\rm fb}^{-1}$ and $300~{\rm fb}^{-1}$ at (a) FCC-ep and (b) ${\rm ILC}\otimes{\rm FCC}$.}\label{fig6}
\end{center}
\end{figure}

In order to identify the isolated lepton or jet, we define the angular separation between particle $i$ and particle $j$ as
\begin{eqnarray}
\label{32}
\Delta R_{ij} = \sqrt{\Delta \phi^2_{ij}+\Delta\eta^2_{ij}}~,
\end{eqnarray}
where $\Delta \phi_{ij}=\phi_{i}-\phi_{j}$ and $\Delta\eta_{ij}=\eta_{i}-\eta_{j}$ with $\phi_{i}$ ($\eta_{i}$) the azimuthal angle (rapidity) of the related lepton or jet.
In the following numerical calculations, we apply the basic acceptance cuts
\begin{eqnarray}
\label{33}
p_{T}^{\ell} > 20~{\rm GeV}  ,  |\eta^{\ell}| < 2.5  , p_{T}^{j} > 20~{\rm GeV} , |\eta^{j}| < 5 \; ,
\end{eqnarray}
\begin{eqnarray}
\label{34}
\min\{\Delta R_{\ell\ell},~\Delta R_{\ell j},~\Delta R_{jj}\} > 0.4 \; .
\end{eqnarray}
The dominant backgrounds in the standard model for the signal process are $e^{-}p \rightarrow e^{-}(\nu_{e})ZZjX$ and $e^{-}p \rightarrow e^{-}(\nu_{e})ZW^{\pm}W^{\mp}jX$ with $Z \rightarrow \mu^{+}\mu^{-}$ and $W^{\pm} \rightarrow \mu^{\pm}\nu_{\mu}$, which are simulated by MadGraph~\cite{Alwall:2014hca}.
To reduce the intermediate $Z$ boson backgrounds, any opposite-sign dimuon pairs with invariant mass close to $M_{Z}$ are rejected
\begin{eqnarray}
\label{35}
\left|M_{\mu^{+}\mu^{-}}-M_{Z}\right| > 20~{\rm GeV} \; .
\end{eqnarray}

With the integraetd luminosity of $100~{\rm fb}^{-1}$ and $300~{\rm fb}^{-1}$, we display the statistical significance as a function of the doubly charged Higgs mass at FCC-ep and ILC$\otimes$FCC in Fig.~\ref{fig6}, where the statistical significance is defined as $S/\sqrt{B}$ with $S$ ($B$) being the signal (background) event numbers after cuts.
It shows that, at FCC-ep, the upper limit of the doubly charged Higgs mass is 400 GeV (418 GeV) with the integrated luminosity of $100~{\rm fb}^{-1}$ ($300~{\rm fb}^{-1}$) for 2$\sigma$ discovery, which is already ruled out by the LHC experiments.
At ILC$\otimes$FCC, with the integrated luminosity of $100~{\rm fb}^{-1}$ ($300~{\rm fb}^{-1}$), the doubly charged Higgs mass can reach 1708 GeV (1818 GeV) at 2$\sigma$ significance, 1622 GeV (1734 GeV) at 3$\sigma$ significance, and 1518 GeV (1630 GeV) at 5$\sigma$ significance.

\subsection{Same-sign diboson decay mode}\label{subsec5.2}

\begin{figure}[!htbp]
\begin{center}
\includegraphics[width=0.40\textwidth]{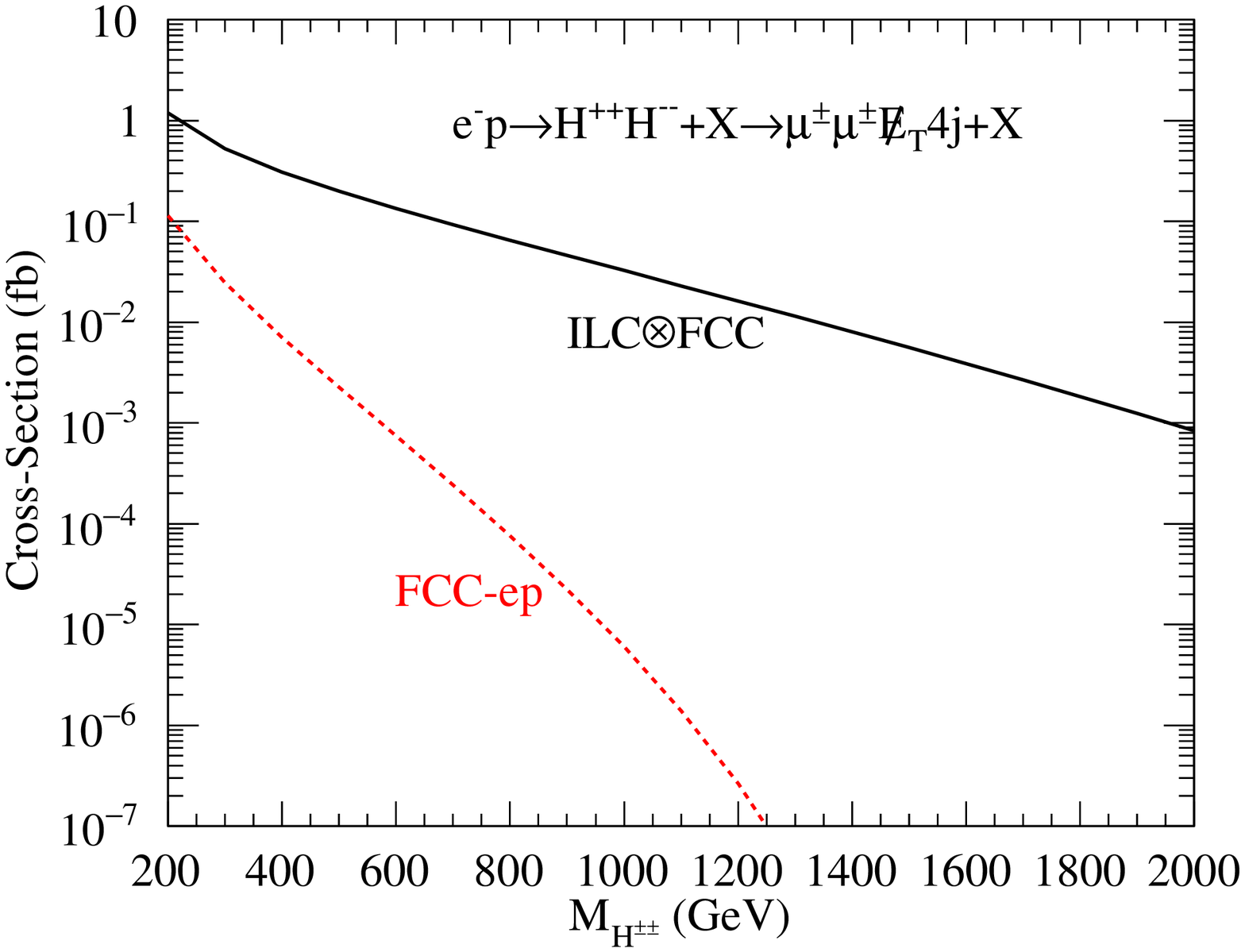}
\caption{ The inclusive production cross sections for $e^{-}p\rightarrow H^{++} H^{--} + X\rightarrow \mu^{\pm}\mu^{\pm}{E\slash}_{T}4j+ X$ at FCC-ep and ${\rm ILC}\otimes{\rm FCC}$ as a function of $M_{H^{\pm\pm}}$.}\label{fig7}
\end{center}
\end{figure}

As mentioned earlier, for large values of $v_{\Delta}$, the doubly charged Higgs decays will be dominated by a pair of same-sign $W$ bosons.
In this case, we focus on investigating the doulby charged Higgs in the following processes:
\begin{eqnarray}
\label{36}
  e^{-} + p \rightarrow e^{-} + H^{++} + H^{--} + j  \; , \quad {\rm with} \quad \left\{
  \begin{array}{ll}
  H^{++} \rightarrow W^{+}W^{+} \; , \;
  H^{--} \rightarrow W^{-}W^{-}  \; ,   \\
  W^{+}W^{+} W^{-}W^{-} \rightarrow \ell^{\pm}\ell^{\pm}{E\slash}_{T} + 4j \; ,
  \end{array} \right.  \nn\\
  e^{-} + p \rightarrow \nu_{e} + H^{++} + H^{--} + j  \; , \quad {\rm with} \quad \left\{
  \begin{array}{ll}
  H^{++} \rightarrow W^{+}W^{+} \; , \;
  H^{--} \rightarrow W^{-}W^{-}  \; ,  \\
  W^{+}W^{+} W^{-}W^{-} \rightarrow \ell^{\pm}\ell^{\pm}{E\slash}_{T} + 4j \; .
  \end{array} \right.
\end{eqnarray}
For the processes in Eq.~(\ref{36}), we demand the two same-sign $W$ bosons in their leptonic decays and the remaining two decay hadronically.
Same as before, we only concentrate on the muon production mode.
Including the decay branching ratios, we display the total cross sections for the inclusive process $e^{-}p\rightarrow H^{++} H^{--} + X\rightarrow \mu^{\pm}\mu^{\pm}{E\slash}_{T}4j+ X$ at FCC-ep and ${\rm ILC}\otimes{\rm FCC}$ in Fig.~\ref{fig7} as a function of $M_{H^{\pm\pm}}$.

\begin{figure}[!htbp]
\begin{center}
\subfigure[]{\label{fig8a}
\includegraphics[width=0.40\textwidth]{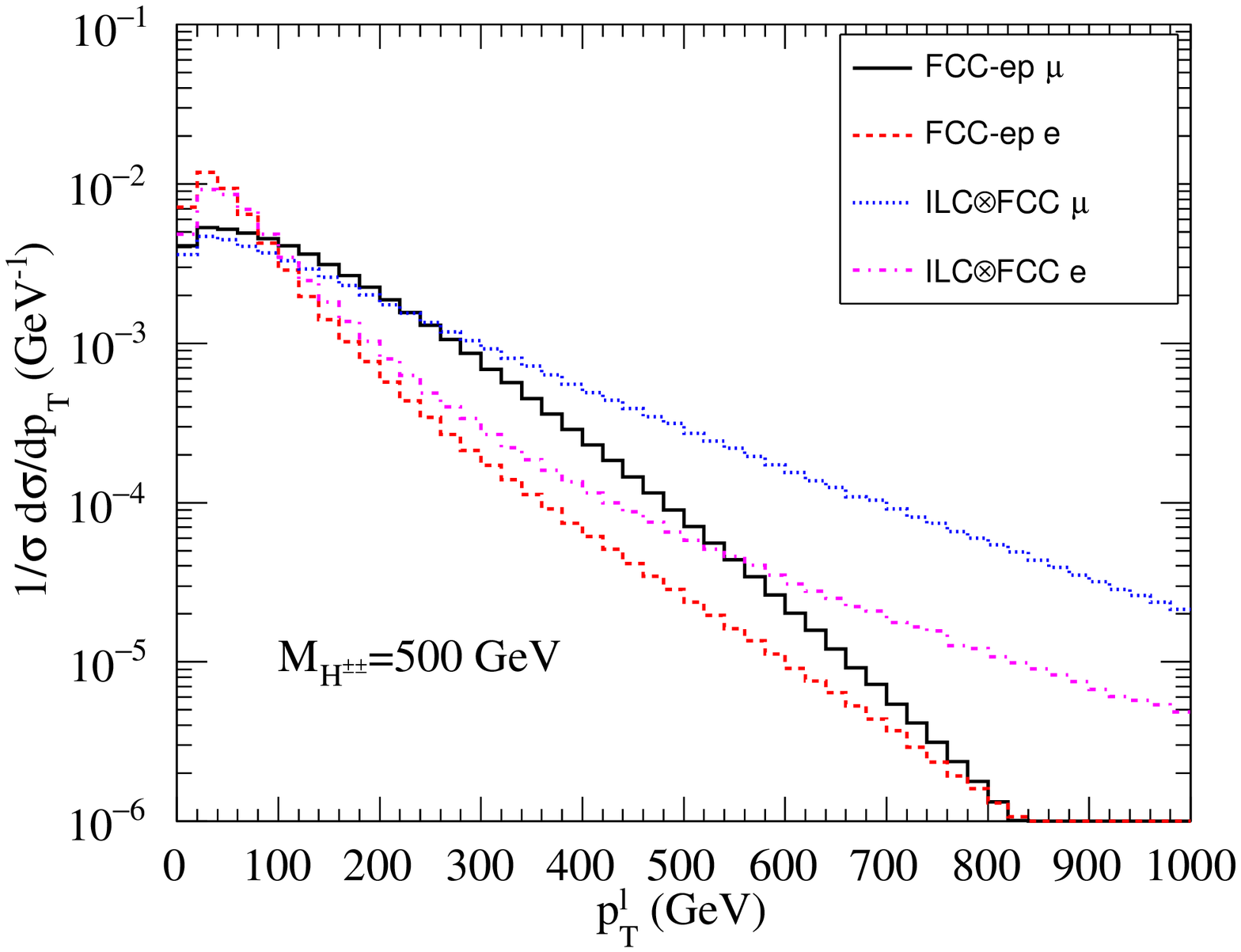} }
\hspace{-0.5cm}~
\subfigure[]{\label{fig8b}
\includegraphics[width=0.40\textwidth]{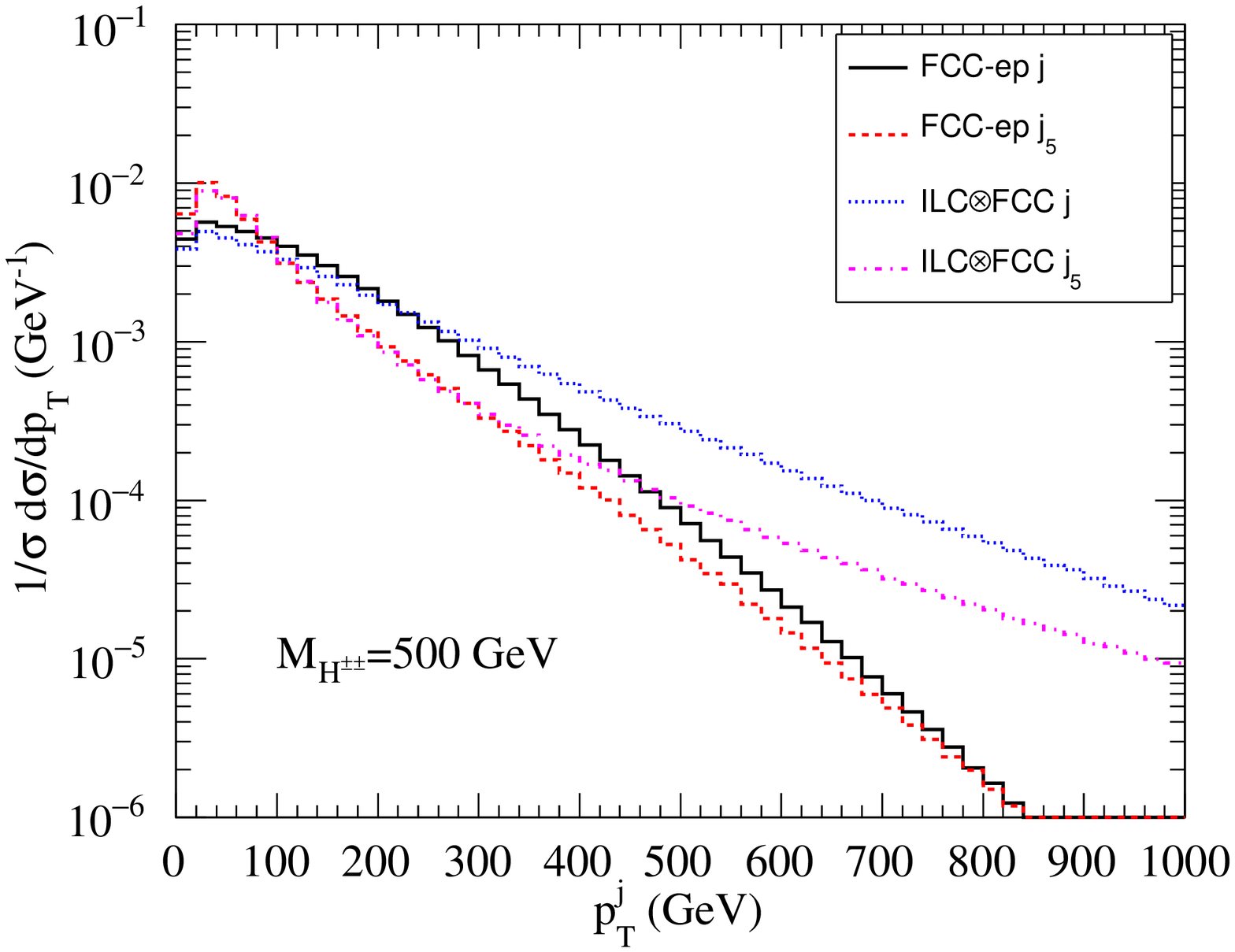} }
\caption{The normalized transverse momentum distributions (a) $1/\sigma{\rm d}\sigma/{\rm d}p^{\mu,e}_{\rm T}$ and (b) $1/\sigma{\rm d}\sigma/{\rm d}p^{j}_{\rm T}$ for $M_{H^{\pm\pm}}=500~{\rm GeV}$ at FCC-ep and ${\rm ILC}\otimes{\rm FCC}$.}\label{fig8}
\end{center}
\end{figure}

After smearing the lepton and jet energies with the help of Eq.~(\ref{31}), we investigate the transverse momentum distributions of the final state particles for the processes in Eq.~(\ref{36}).
The two muons in the final states are labeled as $\mu_{i}$ ($i=1,2$) and are ranked by $p_{T}$ with $p_{T}^{\mu_{1}}>p_{T}^{\mu_{2}}$.
Analogously, the transverse momentum differential distribution of the two muons is defined as $1/\sigma{\rm d}\sigma/{\rm d}p^{\mu}_{\rm T}=1/\sigma({\rm d}\sigma/{\rm d}p^{\mu_{1}}_{\rm T}+{\rm d}\sigma/{\rm d}p^{\mu_{2}}_{\rm T})/2$.
For jets, the four jets $j_{i}$ ($i=1,2,3,4$) that possess an invariant mass closest to $M_{H^{\pm\pm}}$ are selected and are ranked by $p_{T}$ with $p_{T}^{j_{1}}>p_{T}^{j_{2}}>p_{T}^{j_{3}}>p_{T}^{j_{4}}$.
We define the transverse momentum differential distribution of the selected four jets as $1/\sigma{\rm d}\sigma/{\rm d}p^{j}_{\rm T}=1/\sigma({\rm d}\sigma/{\rm d}p^{j_{1}}_{\rm T}+{\rm d}\sigma/{\rm d}p^{j_{2}}_{\rm T}+{\rm d}\sigma/{\rm d}p^{j_{3}}_{\rm T}+{\rm d}\sigma/{\rm d}p^{j_{4}}_{\rm T})/4$.
The remaining jet produced in association with $H^{++}H^{--}$ is denoted by $j_5$.
In Fig.~\ref{fig8}, we plot the normalized transverse momentum distributions $1/\sigma{\rm d}\sigma/{\rm d}p^{\mu,e,j,j_5}_{\rm T}$ for $M_{H^{\pm\pm}}=500~{\rm GeV}$ at FCC-ep and ${\rm ILC}\otimes{\rm FCC}$, respectively.

\begin{figure}[!htbp]
\begin{center}
\subfigure[]{\label{fig9a}
\includegraphics[width=0.40\textwidth]{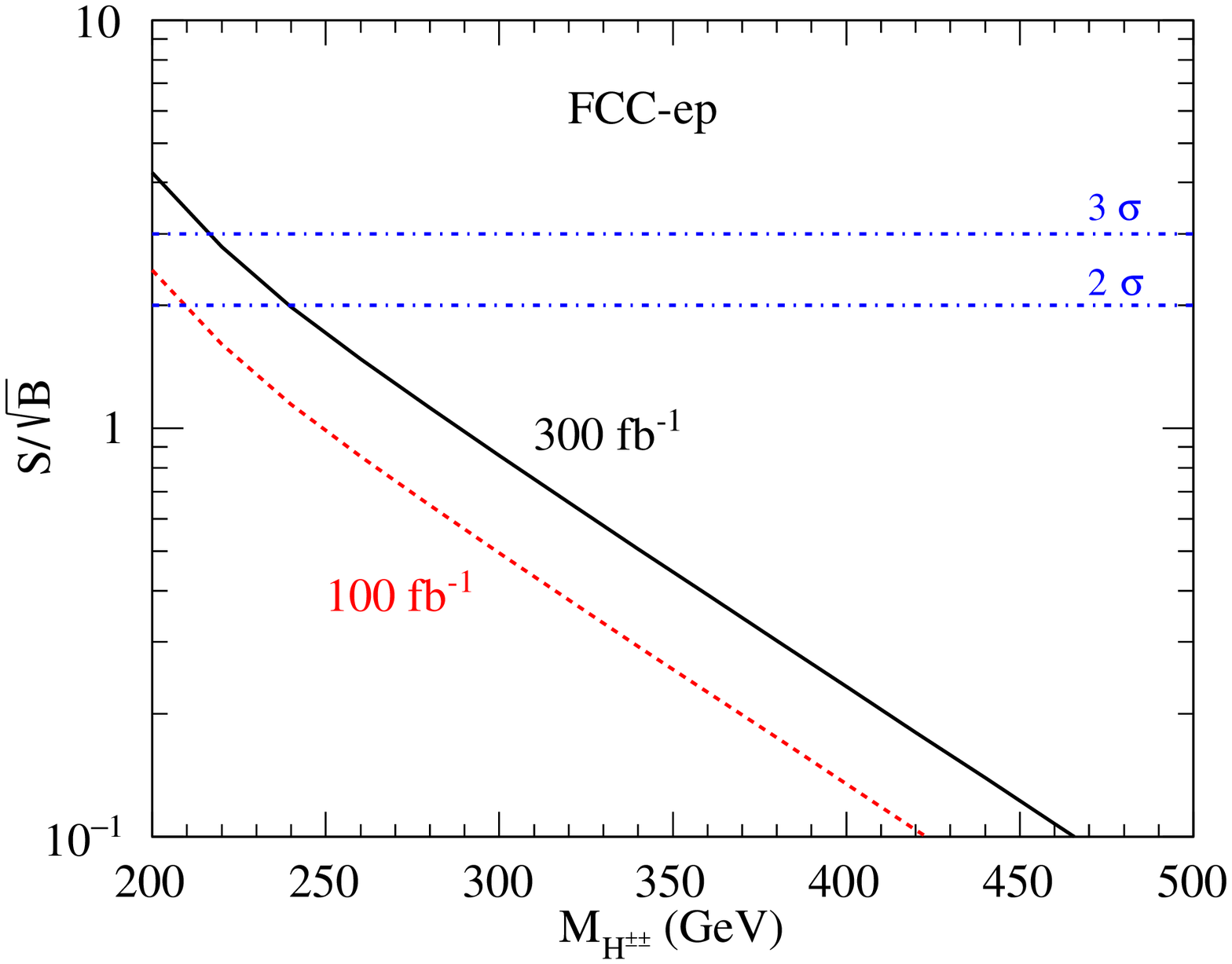} }
\hspace{-0.5cm}~
\subfigure[]{\label{fig9b}
\includegraphics[width=0.40\textwidth]{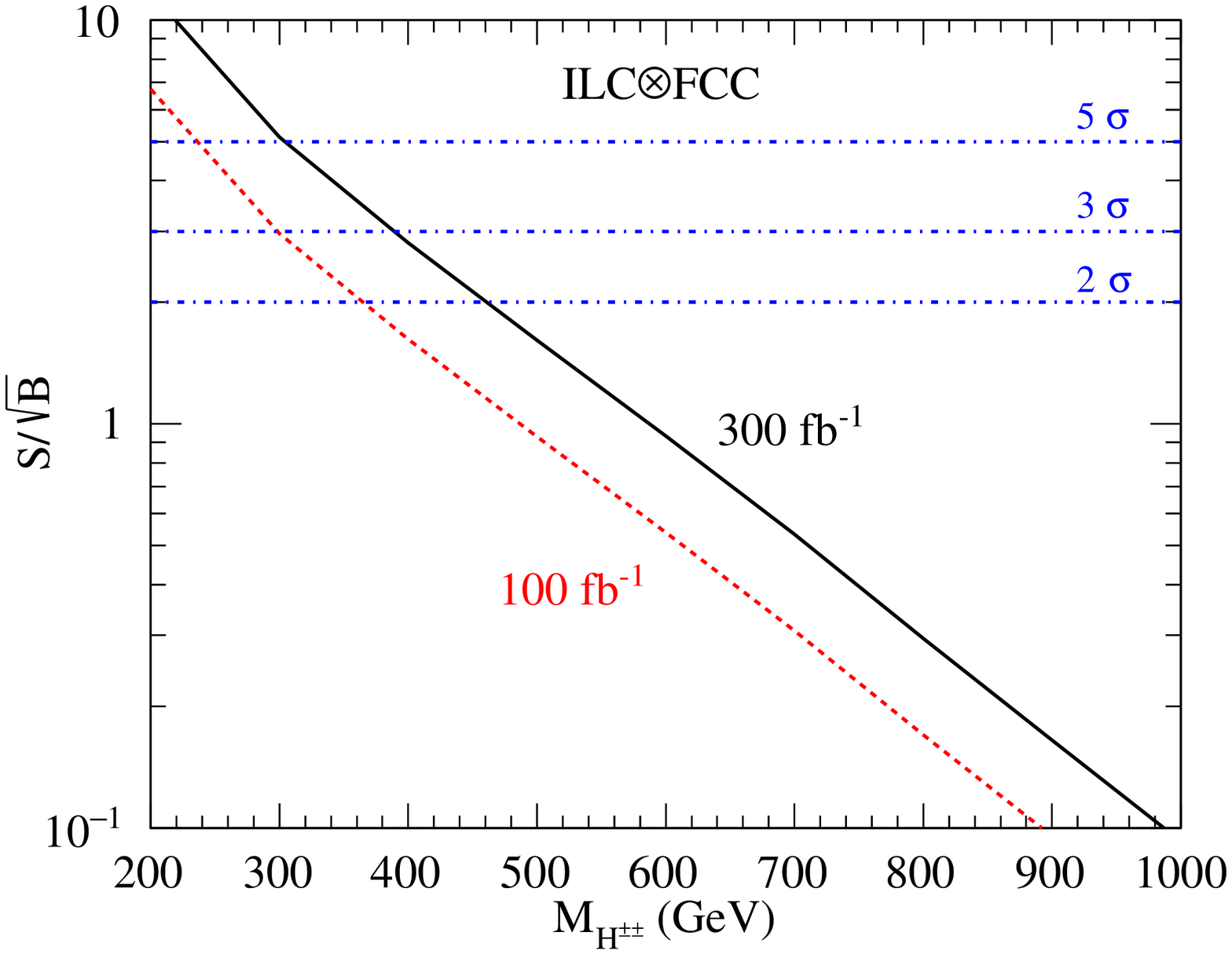} }
\caption{The statistical significance as a function of the doubly charged Higgs mass with the integraetd luminosity of $100~{\rm fb}^{-1}$ and $300~{\rm fb}^{-1}$ at (a) FCC-ep and (b) ${\rm ILC}\otimes{\rm FCC}$.}\label{fig9}
\end{center}
\end{figure}

We start with the following basic cuts
\begin{eqnarray}
\label{37}
p_{T}^{\ell} > 20~{\rm GeV} \; , \quad |\eta^{\ell}| < 2.5  \; , \quad p_{T}^{j} > 20~{\rm GeV}  \; , \quad |\eta^{j}| < 5 \; ,
\end{eqnarray}
\begin{eqnarray}
\label{38}
\min\{\Delta R_{\ell\ell},~\Delta R_{\ell j},~\Delta R_{jj}\} > 0.4  \; , \quad {E\slash}_{T} > 30~{\rm GeV} \; .
\end{eqnarray}
The leading background to the signal is
\begin{eqnarray}
\label{39}
e^-p \rightarrow e^-(\nu_e) t\bar{t}W^{\pm}j
\rightarrow e^-(\nu_e) b\bar{b}W^{+}W^{-}W^{\pm}j
\rightarrow e^-(\nu_e) \mu^{\pm}\mu^{\pm} b\bar{b}jjj\nu_{\mu}\nu_{\mu} \; .
\end{eqnarray}
In order to further purify the signal, the invariant mass of $j_1,j_2,j_3,j_4$ close to $M_{H^{\pm\pm}}$ is required
\begin{eqnarray}
\label{40}
|M_{j_{1}j_{2}j_{3}j_{4}} - M_{H^{\pm\pm}}| < 30~{\rm GeV} \; .
\end{eqnarray}
In Fig.~\ref{fig9}, we display the statistical significance as a function of the doubly charged Higgs mass with the integraetd luminosity of $100~{\rm fb}^{-1}$ and $300~{\rm fb}^{-1}$ at FCC-ep and ILC$\otimes$FCC, respectively.
It is found that, at 2$\sigma$ significance, the upper limit of the doubly charged Higgs mass is 209 GeV (240 GeV) at FCC-ep with the integrated luminosity of $100~{\rm fb}^{-1}$ ($300~{\rm fb}^{-1}$), which is hardly to be probed.
At ILC$\otimes$FCC, with the integrated luminosity of $100~{\rm fb}^{-1}$ ($300~{\rm fb}^{-1}$), the doubly charged Higgs mass up to 364 GeV (461 GeV) can be probed for 2$\sigma$ discovery, while the reach is 300 GeV (392 GeV) for 3$\sigma$ discovery and 238 GeV (308 GeV) for 5$\sigma$ discovery.

\section{Summary}\label{sec6}

In order to explain the tiny neutrino masses, an $SU(2)_L$ Higgs triplet $\Delta$ is introduced to the standard model in the framework of type-II seesaw mechanism.
A typical feature of the type-II seesaw mechanism is that the introduced Higgs triplet can be produced directly through the gauge interactions with the electroweak bosons.
In this work, we explore the pair production of the doubly charged Higgs via vector boson fusion process at future $e^- p$ colliders.
Depending on the size of the Higgs triplet vev, the doubly charged Higgs may decay into a pair of same-sign charged leptons or a pair of same-sign $W$ bosons, we investigate these two decay scenarios in detail.
The total cross sections for the inclusive processes $e^{-}p\rightarrow H^{++} H^{--} + X\rightarrow 2\mu^{+}2\mu^{-}+ X$ and $e^{-}p\rightarrow H^{++} H^{--} + X\rightarrow \mu^{\pm}\mu^{\pm}{E\slash}_{T}4j+ X$ at FCC-ep and ${\rm ILC}\otimes{\rm FCC}$ are predicted.
Furthermore, the transverse momentum distributions of the final state particles for the signal processes are studied.
Finally, we derive the discovery potential of the doubly charged Higgs at future $e^- p$ colliders.
It is found that one can hardly probe the doubly charged Higgs at FCC-ep.
However, at ILC$\otimes$FCC, with the integrated luminosity of $100~{\rm fb}^{-1}$ ($300~{\rm fb}^{-1}$), the doubly charged Higgs mass can reach 1708 GeV (1818 GeV) at 2$\sigma$ significance, 1622 GeV (1734 GeV) at 3$\sigma$ significance, and 1518 GeV (1630 GeV) at 5$\sigma$ significance for dilepton decay mode, while the reach is 364 GeV (461 GeV) at 2$\sigma$ significance, 300 GeV (392 GeV) at 3$\sigma$ significance and 238 GeV (308 GeV) at 5$\sigma$ significance for diboson decay mode.

\section*{Acknowledgments}

The authors would like to thank Profs. Yi Jin, Hong-Lei Li, Shi-Yuan Li and Zong-Guo Si for their helpful discussions.
This work is supported in part by National Natural Science Foundation of China (grant Nos. 11605075, 11635009)
and Natural Science Foundation of Shandong Province (grant No. ZR2017JL006).

\end{document}